\documentclass{article}

\usepackage{arxiv}

\usepackage[utf8]{inputenc} 
\usepackage[T1]{fontenc}    
\usepackage{hyperref}       
\usepackage{url}            
\usepackage{booktabs}       
\usepackage{amsmath, amsthm, amsfonts}
	\DeclareMathOperator*{\argmin}{arg\,min}
	\theoremstyle{theorem}
	\newtheorem{lemma}[]{Lemma}
	\newtheorem{theorem}[]{Theorem}
	\newtheorem{corollary}[]{Corollary}
	
	\newtheorem{remark}{Remark}
	
	\theoremstyle{definition}
	\newtheorem{definition}[]{Definition}
	\newtheorem{example}[]{Example}

\usepackage{nicefrac}       
\usepackage[noend, ruled, linesnumbered, longend]{algorithm2e}
\usepackage{microtype}      
\usepackage{cleveref}       
	\Crefname{invariant}{Invariant}{Invariants}
\usepackage{lipsum}         
\usepackage{graphicx}
\usepackage{natbib}
\usepackage{doi}

\usepackage{tikz}
	\usetikzlibrary{positioning}
	\usetikzlibrary{arrows}
	\usetikzlibrary{calc}
\usepackage{booktabs}
\usepackage{array}   
\newcolumntype{R}{>{$}r<{$}} 
\usepackage{multirow}
\usepackage{longtable} 
\usepackage{pdflscape}
\usepackage{caption, float}
\usepackage{pifont}
\newcommand{\cmark}{\ding{51}}%
\newcommand{\xmark}{\ding{55}}%
\usepackage{siunitx}
\usepackage{todonotes}
\usepackage[super]{nth}

\usepackage{xcolor}
\newcommand{\N}{\mathbb{N}}
\newcommand{\R}{\mathbb{R}}

\newcommand{\incoming}[1]{\delta^{-}(#1)}
\newcommand{\outgoing}[1]{\delta^{+}(#1)}
\newcommand{\bigoh}[1]{\mathcal{O}\left(#1\right)}

\newcommand{\dom}{\preceq_{D}}

\newcommand{\ts}{\textsuperscript}

\newcommand{\kspp}[1]{$#1$-SSP}
\newcommand{\ksppInst}[1]{$\mathcal{I} := (D, s,t, c, #1)$}
\newcommand{\subpath}[3]{#1^{#2 \rightarrow #3}}

\newcommand{\bdassp}{$\text{BDA}_{2\text{SSP}}$}
\newcommand{\ksppBOSP}[1]{\mathcal{I}_{\text{BOSP}}^{#1}}

\title{$K$-Shortest Simple Paths Using Biobjective Path Search}

\newif\ifuniqueAffiliation

\ifuniqueAffiliation 
\author{ \href{https://orcid.org/0000-0000-0000-0000}{\includegraphics[scale=0.06]{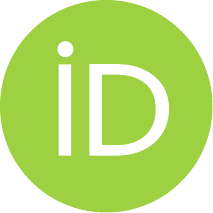}\hspace{1mm}David S.~Hippocampus}\thanks{Use footnote for providing further
		information about author (webpage, alternative
		address)---\emph{not} for acknowledging funding agencies.} \\
	Department of Computer Science\\
	Cranberry-Lemon University\\
	Pittsburgh, PA 15213 \\
	\texttt{hippo@cs.cranberry-lemon.edu} \\
	\And
	\href{https://orcid.org/0000-0000-0000-0000}{\includegraphics[scale=0.06]{orcid.pdf}\hspace{1mm}Elias D.~Striatum} \\
	Department of Electrical Engineering\\
	Mount-Sheikh University\\
	Santa Narimana, Levand \\
	\texttt{stariate@ee.mount-sheikh.edu} \\
}
\else
\usepackage{authblk}

\newbox{\orcid}\sbox{\orcid}{\includegraphics[scale=0.06]{orcid.pdf}} 
\author[1]{%
	\href{https://orcid.org/0000-0002-4197-0893}{\usebox{\orcid}\hspace{1mm}Pedro Maristany de las Casas\thanks{\texttt{maristany@zib.de}}}%
}
\author[2]{%
	\href{https://orcid.org/0000-0003-0681-4585}{\usebox{\orcid}\hspace{1mm}Antonio Sedeño-Noda}%
}
\author[1]{%
	\href{https://orcid.org/0000-0001-7223-9174}{\usebox{\orcid}\hspace{1mm}Ralf Borndörfer}%
}
\author[-]{%
	Max Huneshagen%
}
\affil[1]{Network Optimization Department, Zuse Institute Berlin, Berlin, Germany 14195}
\affil[2]{Facultad de Matematicas, Estadistica e Investigacion Operativa, Universidad de La Laguna, Spain 38200}
\fi

\renewcommand{\headeright}{Preprint}

\hypersetup{
pdftitle={K-Shortest Simple Paths Using Biobjective Path Search},
pdfauthor={P. Maristany de las Casas, A. Sedeño-Noda, R. Borndörder, M. Huneshagen},
pdfkeywords={K-Shortest Simple Paths, Biobjective Search, Second Simple Shortest Path, Dynamic Programming},
}

\begin{document}
\maketitle

\begin{abstract}
	In this paper we introduce a new algorithm for the \emph{$k$-Shortest Simple Paths} (\kspp{k}) problem 
	with an asymptotic running time matching the state of the art from the literature. 
	It is based on a black-box algorithm due to \citet{Roditty12} that solves
	at most $2k$ instances of the \emph{Second Shortest Simple Path} (\kspp{2}) problem without specifying how this is done. 
	We fill this gap using a novel 
	approach:
	we turn the scalar \kspp{2} 
	into instances of the Biobjective Shortest Path problem. Our experiments on grid graphs and on road networks show that the new algorithm is very efficient in practice.
\end{abstract}


\funding{ZIB authors conducted this work within the Research Campus MODAL - Math. Optimization and Data Analysis Laboratories -, funded by the German Federal Ministry of Education and Research (BMBF) (fund number 05M20ZBM).}

\section{Introduction}

Given a directed graph $D = (V,A)$ and a scalar arc cost function $c: A \to \R_{\geq 0}$, we assume paths to be tuples of arcs and define a path's cost as the sum of the cost of its arcs. Then, the optimization problem treated in this paper, the $k$-Shortest Simple Path problem is defined as follows.

\begin{definition}\label{def:k-shortest}
	Given a directed graph $D = (V,A)$, two nodes $s$, $t \in V$, an arc cost function $c: A \to \R_{\geq 0}$, and an integer $k \geq 2$, let $P_{st}$ be the set of simple $s$-$t$-paths in $D$. Assume $P_{st}$ contains at least $k$ paths. The \emph{$k$-Shortest Simple Path} (\kspp{k}) problem is to find a sequence $P=(p_1, p_2, \dots, p_k)$ of pairwise distinct $s$-$t$-paths with $c(p_i) \leq c(p_{i+1})$ for any $i \in \{1, \dotsc, k-1\}$, s.t. there is no path $p \in P_{st} \setminus P$ with $c(p) < c(p_k)$.
	We refer to the tuple $\mathcal{I} := (D, s,t, c, k)$ as a \kspp{k} instance and call $P$ a solution sequence.
\end{definition}

%
\subsection{Literature Overview}
\label{sec:kspp:intro}

\begin{figure}
	\includegraphics[width=\linewidth]{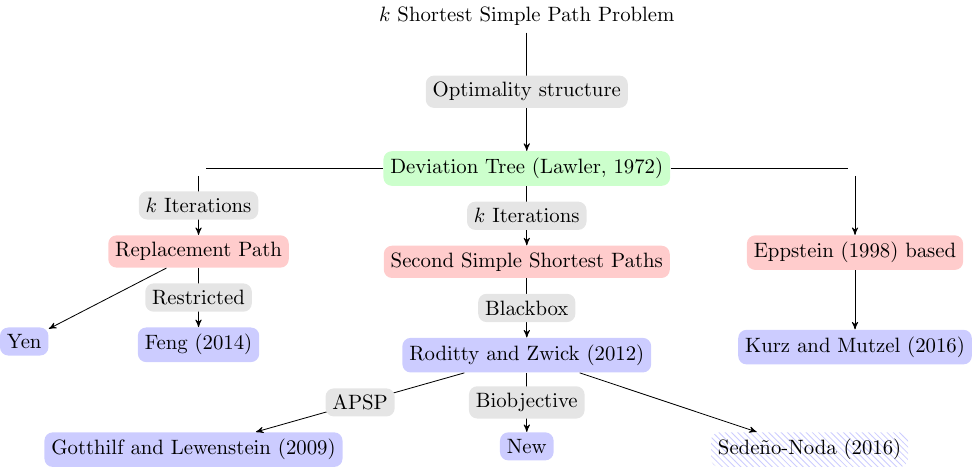}
	\caption{Relevant literature for this paper. The green node represents the optimality structure used by all algorithms to keep track of the solution paths as they are found and to avoid duplicates (see \Cref{sec:optStructure}). Red nodes symbolize solution approaches. Blue nodes refer to \kspp{k} algorithms. If a blue node has solid background, the corresponding algorithm has a state of the art asymptotic running time.}
	\label{fig:kspp:literatureOverview}
\end{figure}

The oldest reference on the \kspp{k} we could find in the literature is the work by \citet{Clarke63}. A detailed and highly recommendable literature survey on this topic is given by \citet{Eppstein2016}.  \Cref{fig:kspp:literatureOverview} gives a visual overview of publications that are relevant for our paper. The figure serves also as an outline for this section. We assume the reader is familiar with basic concepts in \emph{Multicriteria Optimization}; necessary background can be read in e.g., \citep{Ehrgott2005}.

To solve the \kspp{k} problem efficiently, algorithms need to keep track of the $s$-$t$-paths found so far and be able to avoid the generation of duplicates without the need to pairwise compare a new path with every existing path. All relevant \kspp{k} algorithms do so using an \emph{optimality structure} called \emph{deviation tree} first devised by \citet{Lawler72}, to be discussed in \Cref{sec:optStructure}.
%
It is based on the consideration of subpaths.

\begin{definition}
	Given a digraph $D = (V,A)$ and a simple $u$-$w$-path $p$ in $D$ between distinct nodes $u, w \in V$, we denote a subpath of $p$ between nodes $v, v' \in p$ by $\subpath{p}{v}{v'}$. Thereby, if $v = s$ we call $\subpath{p}{v}{v'} = \subpath{p}{s}{v'}$ a \emph{prefix} of $p$ and if $v' = t$ we call $\subpath{p}{v}{v'} = \subpath{p}{v}{t}$ a \emph{suffix} of $p$. 
	For a node $v$ along $p$, we write $v\in p.$
\end{definition}   

Undoubtedly the classical \kspp{k} algorithm is due to \citet{Yen1972}. It performs $k$ iterations and starts with a solution sequence $P = (p_1)$ containing only a shortest $s$-$t$-path $p_1$. In the $i$\ts{th} iteration, $i \in \{1,\dotsc, k\}$, an $i$\ts{th} shortest $s$-$t$-path $p_i$ is considered and the following set of $s$-$t$-paths is computed.

\begin{equation}
	\label{eq:replacementPath}
	\big\{ q \ \big| \  q = \subpath{p_i}{s}{v} \circ \subpath{q}{v}{t}, \, \subpath{q}{v}{t} \text{ shortest $v$-$t$-path in } D \setminus \{\outgoing{v} \cap p_i \}, \, v\in p_i            \big\}.
\end{equation}

The set is a solution to the so called \emph{Replacement Path} (RP) problem. The paths from this set and from all such sets computed in previous iterations are stored in a priority queue of $s$-$t$-paths from which, at the beginning of the $(i+1)$\ts{th} iteration, a $(i+1)$\ts{th} shortest path is extracted and stored in the solution sequence $P$. 
The simple $s$-$t$-path $p_i$ has at most $n$ nodes and thus, \eqref{eq:replacementPath} contains at most $n-1$ elements, each of them requiring a shortest path computation to obtain the suffix $\subpath{q}{v}{t}$. Yen's algorithm solves the RP instances in a straightforward way iterating over the nodes in $p_i$ and solving the corresponding shortest path instances. Using Dijkstra's algorithm \citep{Dijkstra1959} with a Fibnoacci Heap \citep{Fredman1987} for these queries, we obtain a running time for the solution of the RP problem of 

\begin{equation}
	\label{eq:replacementPathOrig}
	T_{RP} := \bigoh{n(n \log{n} + m)}.
\end{equation}

The deviation tree by \citet{Lawler72} (also called \emph{pseudo-tree} in the literature \citep[cf.][]{Martins2003}) is used to ensure that the solution paths for \eqref{eq:replacementPath} computed in every iteration of Yen's algorithm differ from each other without the need to pairwise compare them. Then, Yen's algorithm has an asymptotic running time of
\begin{equation}
	\label{eq:yenRunningTime}
	\bigoh{kT_{RP}}.
\end{equation}

There is a recent alternative \kspp{k} algorithm running in $\bigoh{kT_{RP}}$ that can also be considered the current state of the art and is due to \citet{Kurz2016}. We refer to this algorithm as the \emph{KM algorithm}. Interestingly, the authors achieve this running time without solving the RP problem as a subroutine. Instead, their algorithm can be seen as a generalization of Eppstein's algorithm \citep{Eppstein1998} for the $k$-Shortest Path problem in which the output paths are allowed to contain nodes multiple times. 
Instead of solving One-to-One Shortest Path instances as required in \eqref{eq:replacementPath}, the KM algorithm solves One-to-All Shortest Path instances, hence obtaining a shortest path tree from every search. 
These instances are defined on the reversed input digraph and are rooted at the target node. 
The main idea of the KM algorithm, similar to the idea in \citep{Eppstein1998}, is that $\bigoh{m}$ simple $s$-$t$-paths can be obtained from such a tree using non-tree arcs to create alternative $s$-$t$-paths. 
By doing so, a cycle may be constructed in which case the KM algorithm needs to compute a new shortest path tree.
In addition to its state of the art running time bound, the efficiency of the KM algorithm in practice is immediately apparent: in \emph{well behaved} networks, only few shortest path trees are needed since the swapping of tree arcs and non-tree arcs yield enough simple $s$-$t$-paths. 
Indeed, in the computational experiments conducted in \citep{Kurz2016}, the KM algorithm clearly outperforms the previous state of the art \kspp{k} algorithm by \citet{Feng2014}. This algorithm resembles Yen's algorithm but partitions nodes into three classes, being able to ignore nodes from one of the classes while solving \eqref{eq:replacementPath}. Due to the reduced search space/graph, the One-to-One Shortest Path computations finish faster than in Yen's algorithm.

\subsubsection{Better Asymptotics and Better Computational Performance}


The algorithm by \citet{Gotthilf2009} (\emph{GL algorithm}) improves the best known asymptotic running time for the \kspp{k} problem. It makes use of the All Pairs Shortest Path (APSP) algorithm introduced in \citet{Pettie2004} to achieve an asymptotic running time bound of
$
\bigoh{k(n^2 \log{\log{n}} + nm)}.
$
Here, the term $(n^2 \log{\log{n}} + nm)$ corresponds to the APSP running time bound derived by Pettie, while $k$ APSP instances need to be solved in the GL algorithm. As a brief digression from the main focus 
of the paper, we remark that a new APSP algorithm published in \citet{Orlin22} achieves an asymptotic running time bound of $\bigoh{mn}$ for instances with nonnegative integer arc costs. Using this new algorithm as a subroutine in the GL algorithm, the following result is immediate.
\begin{theorem}[\kspp{k} Running Time for Integer Arc Costs]
	The \kspp{k} problem from \Cref{def:k-shortest} with integer arc costs can be solved in $\bigoh{kmn}$ time.
\end{theorem}
Despite the unbeaten asypmtotic running time bound, the GL algorithm does not perform well in practice. Solving $k$ APSP instances requires too much computational effort.

There are \kspp{k} algorithms whose asymptotic running time bound is worse than \eqref{eq:yenRunningTime} and possibly not even pseudo-polynomial but that perform extremely well in practice. The current state of the art among these algorithms is published in \citet{SedenoNoda2016} and in \citet{Feng2014a}, the latter publication being 
based on the MPS algorithm \citep{MARTINS1999}. Both algorithms are very different from the ones we study here and space limits detain us from discussing  them in more detail.

\subsection{Contribution and Outline}

\Cref{fig:kspp:literatureOverview} shows that there is a third approach to 
the \kspp{k} problem. 
Namely, 
\citet{Roditty2005, Roditty12} show that the \kspp{k} problem can be tackled by solving at most $2k$ instances of the \emph{Second Simple Shortest Path} (\kspp{2}) problem. In their publications, the authors do not specify how the \kspp{2} instances arising as subproblems in their algorithm can be solved efficiently.

%
We design, for the first time, a computationally competitive version of the black box algorithm by Roditty and Zwick. To do so we use a novel algorithm for the \kspp{2} problem. This algorithm is based on a One-to-One version of the recently published Biobjective Dijkstra Algorithm (BDA) \citep{Sedeno19, Casas21, Maristany21}. 

Algorithms that generate (shortest) paths w.r.t. a scalar arc cost function iteratively are known as \emph{ranking algorithms}. These algorithms are sometimes used to solve the \kspp{k} problem \citep{SedenoNoda2016} or to solve Biobjective Shortest Path (BOSP) instances hoping that the generated paths are optimal in the given biobjective setting \citep[cf.][]{MARTINS1999}. Ranking approaches to solve BOSP problems seem nowadays outdated given the improved efficiency reached by recent algorithms \citep{Sedeno19, Ahmadi2021, MaristanydelasCasas2021}. Our approach of solving \kspp{2} instances by defining a BOSP instance and solving it using the BDA turns around the strategies used so far: the \kspp{k} problem, suitable for ranking algorithms, is solved solving $\bigoh{k}$ BOSP instances as subroutines.

In \Cref{sec:optStructure} we describe the \emph{deviation tree}, the optimality structure used throughout the chapter. In \Cref{sec:new2ssp} we discuss our main contribution: a new \kspp{2} algorithm using a biobjective approach. Even if it might sound counter-intuitive to define a biobjective subroutine for an optimization problem with scalar costs, its running time matches the running time bound \eqref{eq:replacementPathOrig}. In \Cref{sec:roditty} we describe the \kspp{k} algorithm by \citet{Roditty12} that solves $\bigoh{k}$ \kspp{2} instances. In the final \Cref{sec:kspp:experiments} we demonstrate the efficiency of our algorithm in practice, benchmarking it against the KM algorithm \citep{Kurz2016}.

\section{Optimality Structure -- Deviation Tree}
\label{sec:optStructure}

Consider a \kspp{k} instance \ksppInst{k}. A (partial) solution sequence $P_{\ell} = (p_1, \dotsc, p_{\ell})$ for $\ell \in [1,k]$ is represented as a \emph{deviation tree} $T_{P_{\ell}}$ \citep[e.g., ][]{Lawler72, Martins2003, Roditty12}. $T_{P_{\ell}}$ is a directed graph, represented as a tree in which a node from the original graph $D$ may appear multiple times. The root node of $T_{P_{\ell}}$ is a copy of the node $s$ in $D$ and every leaf corresponds to a copy of the node $t$. There are $\ell$ leafs and any path from the root to a leaf is in one-to-one correspondence with an $s$-$t$-path in $D$.

\begin{definition}[Deviation Tree.]
	\label{def:kspp:deviations}
	The deviation tree $T_{P_{\ell}}$ is built iteratively. Initially, $T_{P_{\ell}}$ is empty. $p_1$ is added to $T_{P_{\ell}}$ by adding all nodes and edges of $p_1$ to the tree. For any $j \in [2, \ell]$, assume that the previous paths $p_i$, $i \in [1, j)$ have been added to $T_{P_{\ell}}$ already. Assume the longest common prefix of $p_j$ with a path $p_i \in P_{j-1}$ is the $s$-$v$-subpath $\subpath{p_j}{s}{v}$ for a $v \in p_j$. Then, $p_j$ is added to $T_{P_{\ell}}$ by appending the suffix $\subpath{p_j}{v}{t}$ of $p_j$ to the copy of $v$ along $p_i$ in $T_{P_{\ell}}$. 
	\begin{description}
		\item[Parent Path] The path $p_1$ has no parent path. For any path $p_i$, $i \in \{2, \dotsc, \ell\}$, the \emph{parent path} $p$ is the path in $P_{\ell}$ with which $p_i$ shares the longest (w.r.t. the number of arcs) common prefix $\subpath{p_i}{s}{v}$. In case $p$ is not uniquely defined, $p$ is set to be the first path in $P_{\ell}$ with $\subpath{p}{s}{v} = \subpath{p_i}{s}{v}$. If $p$ is the parent path of $p_i$, $p_i$ is a \emph{child path} of $p$.
		\item[Deviation Arc, Deviation Node, Source Node] The path $p_1$ has no \emph{deviation arc}, its \emph{deviation node} is $s$ and it its \emph{source node} is also $s$. For any path $p_i$, $i \in \{2, \dotsc, \ell\}$ the \emph{deviation arc} is the first arc $(v,w)$ along $p_i$ after the common prefix of $p_i$ with its parent path. The node $v$ is called the \emph{deviation node} of $p_i$ and the node $w$ is called the \emph{source node} of $p_i$. For any path $p \in P_{\ell}$ we write $\mathtt{dev}(p)$ and $\mathtt{source}(p)$ to refer to these nodes.
	\end{description}
\end{definition}

Recall that we assume the paths in $P_{\ell}$ to be sorted non-decreasingly according to their costs. Then, the parent path $p$ of any path $q \in P_{\ell}$ is stored before $q$ in $P_{\ell}$ and we have $c(p) \leq c(q)$. Moreover, the inductive nature of $T_{P_{\ell}}$ guarantees that the deviation node of $p$ does not come after the deviation node of $q$.

\begin{example}
	The left hand side of \Cref{fig:kspp:example} shows a \kspp{4} instance. We set $P = (p_1, \dotsc, p_4)$ with $p_1 = ((s,v), (v,t))$, $p_2 = ((s,u), (u,t))$, $p_3=((s,w),(w,t))$, and $p_4=((s,u),(u,v),(v,t))$. The right hand side depicts the deviation tree $T_P$ as defined in \Cref{def:kspp:deviations}. When building $T_P$ iteratively, $p_1$ is added first. Then, the longest common prefix of $p_1$ and $p_2$ is identified to be just the node $s$. Thus, $p_2$ is appended to $s$ in $T_P$. The parent path of $p_2$ is $p_1$, the deviation node is $s$, and the source node is $u$. Adding $p_3$ to $T_P$ leads to the situation in which two paths, namely $p_1$ and $p_2$, share the longest common prefix with $p_3$: the node $s$ only. Hence, the parent path of $p_3$ is set by definition to be $p_1$, the first path in $P$ sharing the longest common prefix with $p_3$. $p_3$'s deviation node is $s$ and its source node is $w$. Finally, the path $p_4$ is added to $T_P$. It shares its prefix $\subpath{p_4}{s}{u}$ of maximum length with $p_2$ and thus its suffix $\subpath{p_4}{u}{t} = ((u,v),(v,t))$ is appended in $T_P$ to the copy of the node $u$ along $p_2$ in $T_P$. The deviation node of $p_4$ is $u$ and its source node is $v$.
	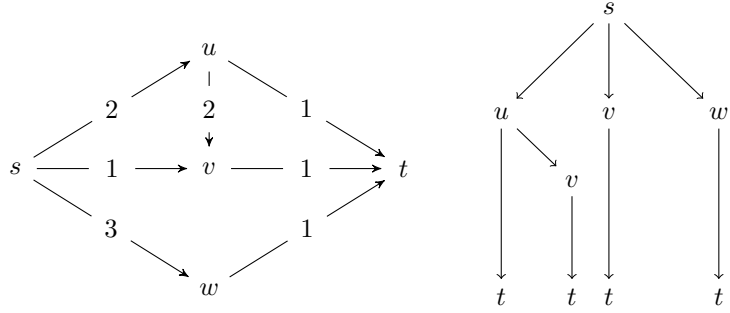
\begin{figure}
		\centering
		\begin{tikzpicture}[every node/.style={draw=none, circle}, >=stealth']
			\node[] at (0, 0) (s) {$s$};
			\node[right=2cm of s] (v) {$v$};
			\node[above=1cm of v] (u) {$u$};
			\node[below=1cm of v] (w) {$w$};
			\node[right=2cm of v] (t) {$t$};
			
			\draw [->] (s) -- (u) node [midway, fill=white] {$2$};
			\draw [->] (s) -- (v) node [midway, fill=white] {$1$};
			\draw [->] (s) -- (w) node [midway, fill=white] {$3$};
			\draw [->] (u) -- (v) node [midway, fill=white] {$2$};
			\draw [->] (u) -- (t) node [midway, fill=white] {$1$};
			\draw [->] (v) -- (t) node [midway, fill=white] {$1$};
			\draw [->] (w) -- (t) node [midway, fill=white] {$1$};
		\end{tikzpicture}
		\qquad
		\begin{tikzpicture}
			\node[] at (0, 0) (sTree) {$s$};
			\node[below left= 1cm and 1cm of sTree] (uTree) {$u$};
			\node[below = 2cm of uTree] (tTree1) {$t$};
			\draw [->] (sTree) -- (uTree);
			\draw [->] (uTree) -- (tTree1);
			
			\node[below = 1cm of sTree] (vTree) {$v$};
			\draw (tTree1 -| vTree) node (tTree2) {$t$};
			\draw [->] (sTree) -- (vTree);
			\draw [->] (vTree) -- (tTree2);
			
			\node[below right= 1cm and 1cm of sTree] (wTree) {$w$};
			\draw (tTree2 -| wTree) node (tTree3) {$t$};                        
			\draw [->] (sTree) -- (wTree);
			\draw [->] (wTree) -- (tTree3);
			
			\node[below right = .5cm and .5cm of uTree] (vTree2) {$v$};
			\draw (tTree1 -| vTree2) node (tTree4) {$t$};
			
			\draw [->] (uTree) -- (vTree2);
			\draw [->] (vTree2) -- (tTree4);
		\end{tikzpicture}
		\caption{Left: Input graph $D$ and arc costs for a \kspp{k} instance. Right: Pseudo-Tree $T_P$ corresponding to the \kspp{k} instance defined on the left with $k=4$.}\label{fig:kspp:example}
	\end{figure}
\end{example}


%
\section{Second Shortest Simple Path Problem}\label{sec:new2ssp}

We introduce a new \kspp{2} algorithm. Assuming that a shortest path $p$ is known, we define a biobjective arc cost function $\gamma_p : A \to \R^2$ depending on $p$ that allows us to find a second shortest path as the first or the second (in lexicographic order w.r.t. $\gamma$) efficient solution of a One-to-One Biobjective Shortest Path (BOSP) instance associated with $p$.

\begin{definition}
	\label{def:kspp:BOSPinstance}
	Consider a digraph $D=(V,A)$, nodes $s, \, t \in V$, and an arc cost function $c : A \to \R_{\geq0}$. $\mathcal{I} = (D,s,t,c)$ is an instance of the classical One-to-One Shortest Path problem. Let $p$ be a shortest $s$-$t$-path in $D$. For every arc $a \in A$ define two dimensional costs $\gamma_p(a) \in \R_{\geq 0}^2$ setting $\gamma_{p,1}(a) = c(a)$ and $\gamma_{p,2}(a) = 1$ if $a \in p$. Otherwise, set $\gamma_{p,2}(a) = 0$. The BOSP instance $\mathcal{I}_{\text{BOSP}}^p = (D, s, t, \gamma_p)$ is the \emph{BOSP instance associated with $\mathcal{I}$ and $p$}.
\end{definition}

Initially, using a biobjective subroutine in an optimization problem with scalar cost sounds counter-intuitive. The reasons are intractability of the Biobjective Shortest Path problem \citep[cf.][]{Hansen1980} and the consequent time and memory demands.
However, the $\gamma_p$ function defined in \Cref{def:kspp:BOSPinstance} is such 
that at most $n-1$ $s$-$t$-paths are optimal (cf. \Cref{lemma:kspp:boundedCardinality}) making the $\ksppBOSP{p}$ instances tractable. Moreover, the $\{0$,$1\}$ second component of $\gamma_p$ plays an essential role to circumvent the issue of second shortest paths not adhering to 
the subpath-optimality principle as explained in the following example.

\begin{example}
	Consider the $\ksppBOSP{p}$ instance defined in \Cref{fig:brokenSubpathOpti} w.r.t. the shortest $s$-$t$-path $p$ in that instance. The shown graph contains two $s$-$v_4$-paths:
	\begin{equation*}
		\begin{split}
			q = ((s,v_1), (v_1,v_4)) \text{ with } c(q) = \gamma_{p,1}(q) = 2 \\
			r = ((s,v_1), (v_1,v_2), (v_2,v_3), (v_3,v_4)) \text{ with } c(r) = \gamma_{p,1}(r) = 1
		\end{split}
	\end{equation*}
	A decision made based on these costs favors the path $r$ since $c(r) < c(q)$. However, the extension of $r$ towards $v_2$ produces a cycle, causing any expansion of $r$ to be an invalid candidate for a second shortest simple $s$-$t$-path. Thus, when comparing $q$ and $r$ both paths need to be recognized as \emph{promising} candidates. Since $q$ shares only one arc with $p$ before it deviates, we have $\gamma_{p,2}(q) = 1$. For the same reason, we have $\gamma_{p,2}(r) = 3$. Thus, $\gamma_p(q) = (2,1)$ and $\gamma_p(r) = (1,3)$ and both paths are efficient/optimal $s$-$v_4$-paths in our biobjective setting.
	
	Note that $v_2$ is already visited by $r$'s subpath $\subpath{r}{s}{v_2}$ with cost $\gamma_p(\subpath{r}{s}{v_2}) = (0,2)$. After expanding $q$ and $r$ along the arc $(v_4, v_2)$, we have $\gamma_p(q \circ (v_4, v_2)) = (4,1)$ and $\gamma_p(r \circ (v_4, v_2)) = (3,3)$ and we see that $r$'s expansion is dominated by $\subpath{r}{s}{v_2}$ and thus can be discarded. $q$'s expansion on the other side is not dominated and thus, the \emph{bad} $s$-$v_4$-path w.r.t. the original cost function $c$ is kept to build a simple second shortest $s$-$t$-path. 
	\begin{figure}
		\centering
		\begin{tikzpicture}[scale=.3]
			\tikzstyle{arc}=[->]
			\tikzstyle{nodo}=[draw]
			\node (s) at (0,0) [] {\footnotesize{$s$}};
			\node[right=2cm of s] (1) {\footnotesize{$v_1$}};
			\node[right=2cm of 1] (2) {\footnotesize{$v_2$}};
			\node[right=2cm of 2] (3) {\footnotesize{$v_3$}};
			\node[right=2cm of 3] (t) {\footnotesize{$t$}};
			
			\node[above=1.5cm of 2] (4) {\footnotesize{$v_4$}};
			
			\draw (s) -- (1) [arc,red] node [midway,fill=white, sloped] {\footnotesize{$(0,1)$}};
			\draw (1) -- (2) [arc,red] node [midway,fill=white,sloped] {\footnotesize{$(0,1)$}};
			\draw (2) -- (3) [arc,red] node [midway,fill=white,sloped] {\footnotesize{$(0,1)$}};
			\draw (3) -- (t) [arc] node [midway,fill=white,sloped] {\footnotesize{$(0,1)$}};
			
			\draw (1) -- (4) [arc] node [midway, fill=white,sloped] {\footnotesize{$(2,0)$}};
			\draw (3) -- (4) [arc,red] node [midway,fill=white,sloped] {\footnotesize{$(1,0)$}};
			\draw (4) -- (2) [arc,red] node [midway,fill=white,sloped] {\footnotesize{$(2,0)$}};
			
			\draw[bend right, arc] (2) to node [fill=white] {\footnotesize{$(2,0)$}} (t);
		\end{tikzpicture}
		\caption{$\ksppBOSP{p}$ instance for the shortest path $p=((s,v_1), (v_1, v_2), (v_2, v_3), (v_3,t))$. The red arcs show a $s$-$v_2$-path that is not simple. Its $s$-$v_4$-subpath is the shortest $s$-$v_4$-path w.r.t. the original arc costs $c = \gamma_1$.}
		\label{fig:brokenSubpathOpti}
	\end{figure}
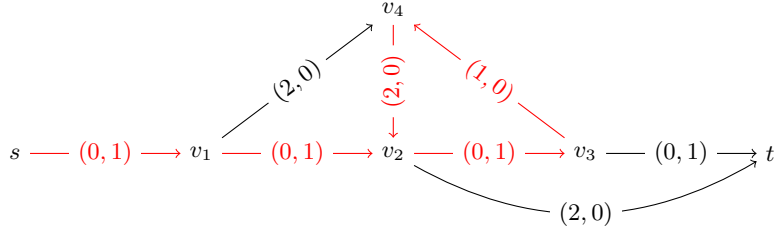
\end{example}

The $\gamma$ arc cost function not only elevates the status of paths with suboptimal subpaths w.r.t. $c$ to become efficient paths in the $\ksppBOSP{}$ instances. We can additionally use a technique called \emph{dimensionality reduction} that allows dominance tests in biobjective optimization problems to be done in constant time. Since, as in the last example, paths that are not simple will turn out to be dominated, we manage to detect cycles without hashing a path's nodes.

The BOSP algorithm to solve the $\ksppBOSP{}$ instances must be chosen carefully to obtain a competitive running time bound for our \kspp{k} algorithm in \Cref{sec:roditty}.

\subsection{Biobjective Dijkstra Algorithm}
The \emph{One-to-One Biobjective Dijkstra Algorithm} (BDA) \citep{Sedeno19, Casas21, Maristany21} is a BOSP algorithm that features the currently best theoretical output sensitive running time bound known for this problem. As the name suggests, it proceeds similarly to Dijkstra's algorithm \citep{Dijkstra1959} for the classical Shortest Path problem. It uses a priority queue in which paths are sorted in lexicographically (lex.) nondecreasing order w.r.t. $\gamma$. While in Dijkstra's algorithm the queue only needs to store the cheapest known $s$-$v$-path for every node $v \in V$, BOSP algorithms prior to the BDA needed to be able to handle multiple $s$-$v$-paths. All explored $s$-$v$-paths that are not dominated by and not cost-equivalent to already stored $s$-$v$-paths need to be stored; here, a path $p$ is said to dominate a path $q$ if $\gamma_i(p) \leq \gamma_i(q)$ for $i \in \{1,2\}$ and one of the inequalities is strict. After an $s$-$v$-path for some $v\in V$ is extracted from the queue, it is stored in the list of optimal $s$-$v$-paths $P^*_{sv}$ and propagated along the outgoing arcs of $v$ to obtain new candidate $s$-$w$-paths, $(v,w) \in \outgoing{v}$. In the biobjective optimization literature, and also hereinafter, optimal solutions are called \emph{efficient}. The BDA returns a \emph{minimum complete set of efficient paths}. This means that if for a non-dominated cost vector $c$ there are multiple efficient $s$-$t$-paths $p$ with $\gamma(p) = c$, the BDA returns just one of these paths.

The main idea in the design of the BDA is that biobjective shortest paths adhere to a dynamic programming principle: the efficient $s$-$v$-paths can be build out of the optimal $s$-$u$-paths for $u \in \incoming{v}$. Exploiting this principle, the BDA manages to store just the lex-smallest non-dominated $s$-$v$-path in its queue. When this path is extracted from the queue and stored in the corresponding list $P^*_{sv}$ of efficient paths, it rebuilds the next candidate $s$-$v$-path looking at the efficient $s$-$u$-paths in $P^*_{su}$ for $(u,v) \in \incoming{v}$. Thanks to this idea, we obtain the following running time and space consumption bound:

\begin{theorem}
	\label{thm:bospRunningTimeAndSpace}
	Let $\mathcal{I} = (D, s, t, \gamma)$ be a BOSP instance and set $N_{\max} := \max_{v \in V} |P^*_{sv}|$ and $N = \sum_{v \in V} |P^*_{sv}|$. The BDA runs in
		$\bigoh{N \log{n} + N_{\max}m}$ time
	and uses
		$\bigoh{N + n + m}$
	space.
\end{theorem}

For further details regarding the (One-to-One) BDA, we refer to the original publications \citep{Sedeno19, Maristany21}. A detailed discussion of the running time and space consumption bounds for the BDA and other BOSP algorithms can be found in \citet{Casas21}.
\subsection{Second Simple Shortest Paths Using the BDA}
We formulate the following main result in this section.

\begin{theorem}
	\label{thm:kssp:correctness}
	Consider a shortest path problem $\mathcal{I} = (D,s,t,c)$,  
	let $p$ be a shortest $s$-$t$-path w.r.t. $c$ and assume it has $\ell$ arcs. A lexicographically smallest (w.r.t. $\gamma_p$) efficient $s$-$t$-path $q$ with $\gamma_{p,2}(q) < \ell$ in the BOSP instance $\mathcal{I}_{\text{BOSP}}^{p}$ is a second shortest simple $s$-$t$-path in $D$ w.r.t. the original costs $c$.
\end{theorem}

\begin{proof}
	We assume that $\mathcal{I}_{\text{BOSP}}^{p}$ is solved using the BDA. Every efficient path in this instance is a simple path or cost-equivalent to a simple path since $\gamma_p$ is a non-negative function. Additionally, efficient paths containing a loop are neither made permanent nor further expanded by the BDA since the algorithm uses the reflexive $\dom$-operator. As a consequence, every path $q$ that is made permanent fulfills $\gamma_{p,2}(q) \leq \ell$ and the only possibly extracted path with $\gamma_{p,2}(q) = \ell$ is $p$ with costs $\gamma_p(p) = (c(p), \ell)$. 
	
	Since $p$ is a shortest $s$-$t$-path, any extracted $s$-$t$-path $q \neq p$ fulfills $\gamma_{p,1}(q) \geq \gamma_{p,1}(p)$. Thus, if $q$ is made permanent before $p$, it is lex-smaller than $p$ and we must have $\gamma_{p,1}(q) = \gamma_{p,1}(p)$ and $\gamma_{p,2}(q) < \gamma_{p,2}(p)$. The second inequality implies that $q$ and $p$ are distinct paths and we thus can stop the execution of the BDA and return $q$ as a second shortest simple $s$-$t$-path. In this case, $p$ and $q$ are cost-equivalent w.r.t. $c$.
	
	Assume $p$ is permanent already and $q$ is the next $s$-$t$-path extracted from the BDA's priority queue. We must have $\gamma_{p,1}(p) < \gamma_{p,1}(q)$ (see last paragraph). Recall that the BDA finds a minimum complete set of efficient paths for $\mathcal{I}_{\text{BOSP}}^{p}$. Moreover, as already noted, paths are extracted from the algorithm's priority queue in \mbox{lex.} nondecreasing order. Thus, since efficient paths are simple, we conclude that there cannot exist a simple $s$-$t$-path $q'$ with $\gamma_{p,1}(p) \leq \gamma_{p,1}(q') < \gamma_{p,1}(q)$ that is not found by the BDA. Since the $\gamma_{p,1}$ costs are equivalent to the original $c$ costs of the paths in $D$, we obtain that $q$ is a second shortest simple $s$-$t$-path.
\end{proof}

The shortest path $p$ is the only simple path in $\mathcal{I}_{\text{BOSP}}^{p}$ with $\gamma_2(p) = \ell$. Since we want to find an efficient $s$-$t$-path $q$ with $\gamma_{p,2}(q) < \ell$, the BDA can stop after at most two $s$-$t$-paths are extracted from the priority queue: the path $p$ that is efficient iff $\gamma_{p,1}(p) < \gamma_{p,1}(q)$ and $q$ itself. Using this stopping criterion, we define the following modified version of the BDA as our new \kspp{2} algorithm.

\begin{definition}[\bdassp{}]
	\label{def:kspp:bdassp}
	The \bdassp{} is a \kspp{2} algorithm. It modifies the BDA as follows.     
	\begin{description}
		\item[Input] In addition to a BOSP instance, the input of the \bdassp{} contains a non-negative integer $\ell$.
		\item[Stopping Condition] The \bdassp{} stops whenever an efficient $s$-$t$-path $q$ with $\gamma_{p,2}(q) < \ell$ is extracted from the priority queue or when the priority queue is empty at the beginning of an iteration.
		\item[Output] Instead of a minimum complete set of efficient $s$-$t$-paths, the new \bdassp{} returns the suffix $\subpath{q}{v}{t}$ of the first efficient $s$-$t$-path $q$ with $\gamma_{p,2}(q) < \ell$ that it finds. Here, $v$ is the node after which $p$ and $q$ deviate for the first time. If such a path $q$ does not exist, the \bdassp{} returns a dummy path if such a path does not exist.
	\end{description}
\end{definition}

\subsection{Asymptotic Running Time and Memory Consumption}

The following is a general statement that holds for any biobjective optimization problem \citep[cf.][]{Sudhoff23}.

\begin{lemma}
	\label{lemma:kspp:boundedCardinality}
	Let $\mathcal{X}$ be the set of feasible solutions of a biobjective optimization problem and $f: \mathcal{X} \to \R_{\geq 0} \times \N$ the associated cost function. The cardinality of a minimum complete set of efficient solutions is bounded by the size of the set $\{f_2(x)\ | \ x \in \mathcal{X}\}$. 
\end{lemma}

\begin{proof}
	Assume for a value $y_2 \in \{f_2(x)\ | \ x \in \mathcal{X}\}$ there are two efficient solutions $x$, $x'$ in a minimum complete set. If $f_1(x) \neq f_1(x')$, then the solution with the smaller $f_1$ value (weakly) dominates the other. If $f_1(x) = f_1(x')$, both solutions are cost-equivalent and, by definition, no minimum complete set contains both.
\end{proof}

Our setting in this section assumes a shortest $s$-$t$-path $p$ with $\ell$ arcs to be given and we look for a second shortest $s$-$t$-path in the same graph. In the first paragraph of the proof of \Cref{thm:kssp:correctness} we derived that any efficient path $q$ for the instance $\mathcal{I}_{\text{BOSP}}^p$ fulfills $\gamma_{p,2}(q) \leq \ell$. \Cref{lemma:kspp:boundedCardinality} applied to $\mathcal{I}_{\text{BOSP}}^{p}$ implies that every minimum complete set of efficient $s$-$v$-paths, $v \in V$, has cardinality at most $\ell$.

For the set of efficient $s$-$t$-paths computed by the \bdassp{} we even now that it contains at most two paths at the end of the algorithm. 
Sadly, we cannot mirror this fact in the running time bound for the \bdassp{}.
As explained in \citet{Boekler18} and \cite{MaristanydelasCasas2021} for a One-to-One BOSP instance, a minimum complete set of efficient $s$-$t$-paths can contain less paths than the number of efficient $s$-$v$-paths calculated for an intermediate node $v$. 
Thus, even though it calculates at most two $s$-$t$-paths, the \bdassp{} may compute $\ell-1$ (not $\ell$ because $s$-$t$-paths are not propagated) $s$-$v$-paths for an intermediate node $v$. Using the running time bound of the BDA described in \Cref{thm:bospRunningTimeAndSpace}, we obtain the following result.

\begin{theorem}
	\label{thm:kspp:bdasspRunningTime}
	The \bdassp{} solves a \kspp{2} instance \ksppInst{2} in time
	\begin{equation}
		\label{eq:bdassp:time}
		\bigoh{n\ell \log{n} + \ell m} = \bigoh{\ell(n \log{n} + m)} \in \bigoh{n(n \log{n} + m)}.
	\end{equation}
\end{theorem}

Based on the memory consumption derived for the BDA in \Cref{thm:bospRunningTimeAndSpace}, we conclude this section stating the space consumption bound of the \bdassp{}.

\begin{theorem}
	\label{thm:kspp:bdasspSpace}
	The \bdassp{} uses
	\begin{equation}
		\label{eq:bdassp:space}
		\bigoh{n\ell + n + m} \in \bigoh{n^2 + n + m} \in \bigoh{n^2 + m}
	\end{equation}
	memory.
\end{theorem}

\begin{proof}
	The original version of the BDA uses $\bigoh{N + n + m}$ space where $N = \sum_{v\in V} N_v$ and $N_v$ is the number of efficient $s$-$v$-paths calculated by the algorithm. We have $N \leq nN_{\max}$ with $N_{\max} = \max_{v\in V} N_v$ and in our \bdassp{} scenario $N_{\max} < \ell$ as discussed already. The modifications defined in \Cref{def:kspp:bdassp} to the original BDA to obtain the \bdassp{} do not have any further impact on the space consumption.
\end{proof}

\subsection{Properties of the Second Shortest Simple Path Problem}

In this section we discuss three structural properties of the \kspp{2} problem. They are helpful for the description of our new \kspp{k} algorithm introduced in \Cref{sec:roditty}. Note that whenever we remove a path $p$ from a given digraph $D$, we write $D \setminus p$ and we delete the nodes and the arcs of $p$ from $D$. 

The following easy statement is essential for the understanding of the remainder of the chapter. The proof follows directly from \Cref{def:kspp:deviations}.

\begin{lemma}
	\label{lemma:secondShortestIsChildOfShortest}
	Consider a \kspp{2} instance and let $P = (p_1, p_2)$ be a solution sequence with associated deviation tree $T_P$. Then, $p_1$ is the parent path of $p_2$.
\end{lemma}

\begin{lemma}
	\label{lemma:kssp:secondShortestShortest}
	Consider a \kspp{2} instance \ksppInst{2} and let $P = (p_1, p_2)$ be a solution sequence. Assume that $(v,w)$ is $p_2$'s deviation arc which is well defined due to \Cref{lemma:secondShortestIsChildOfShortest}. The suffix path $\subpath{p_2}{w}{t}$ is a shortest $w$-$t$-path in the digraph $D \setminus \subpath{p_2}{s}{v}$.
\end{lemma}

\begin{proof}
	We can write $p_2 = \subpath{p_1}{s}{v} \circ (v,w) \circ \subpath{p_2}{w}{t}$. If a $w$-$t$-path $\subpath{q}{w}{t}$ with $c(\subpath{q}{w}{t}) < c(\subpath{p_2}{w}{t})$ exists in $D \setminus \subpath{p_2}{s}{v}$ we have $c(p_2) > c(\subpath{p_1}{s}{v} \circ (v,w) \circ \subpath{q}{w}{t})$ and $p_2$ would not be a second shortest $s$-$t$-path in $D$.
\end{proof}

As a consequence of the last lemma, we formulate the following result.

\begin{corollary}
	\label{coro:kssp:secondShortest}
	A second simple shortest path is given by
	\begin{align}
		\label{eq:secondShortest}
		\begin{split}
			p_2 := \argmin \Big\{  c(q) \ \Big| \   &q = \subpath{p_1}{s}{v} \circ (v,w) \circ \subpath{q}{w}{t}, \,\\ 
			&\subpath{q}{w}{t} \text{ shortest } \text{$w$-$t$-path in } D \setminus \subpath{p_1}{s}{v}, \,\\ 
			&(v,w) \in \outgoing{v}\setminus p_1, \, v \in p_1 \Big\}.
		\end{split}
	\end{align}
\end{corollary}

We observe that the solution $p_2$ in \eqref{eq:secondShortest} is contained in the solution set \eqref{eq:replacementPath} of a Replacement Path (RP) instance. Thus, in a worst case scenario, the \bdassp{} needs to do as much effort as an RP algorithm to find $p_2$ in the set \eqref{eq:replacementPath} of RP solutions. This intuition is formally mirrored in \citet[Theorem 1.1]{Williams2018}. The result states that while currently having the same complexity, a truly subcubic \kspp{2} algorithm implies a truly subcubic RP algorithm. I.e., it is unlikely to design an algorithm solving \kspp{2} instances faster than RP instances in the worst case.

However, for practical purposes the fact that the solution $p_2$ from \eqref{eq:secondShortest} is included in the set \eqref{eq:replacementPath} unveils the strength of the \bdassp{} as a \kspp{2} algorithm: stopping after at most two paths reach the target node $t$ reduces the number of iterations in comparison to the need to solve $\bigoh{n}$ One-to-One Shortest Path instances to calculate \eqref{eq:replacementPath}.

\section{K-SPP Algorithm by Roditty and Zwick using the BDA}
\label{sec:roditty}

\citet{Roditty12} discuss a black box algorithm for the \kspp{k} problem. It is a \emph{black box} algorithm because the authors do not specify how to solve the key subroutine in their algorithm: the computation of a second-shortest simple path. Moreover, they do not implement their algorithm in the paper. In this section we fill the gap using the \bdassp{}.

The algorithm presented in \citep{Roditty12} performs $2k$ computations of a second shortest simple path to solve a \kspp{k} instance \ksppInst{k}. It fills the solution sequence $P$ iteratively. In our exposition we assume that at any stage, the deviation tree $T_P$ associated with $P$ exists implicitly. In particular, this allows us to use the notions from \Cref{def:kspp:deviations}. We discuss this in detail in \Cref{rem:kssp:parentPathsInAlgo}. The pseudocode for our new algorithm is in \Cref{algo:ksppNew}.

\begin{remark}[Source nodes in $\ksppBOSP{}$ instances]
	\label{kspp:rem:SourceNodes}
	Given an $\ell$\ts{th} shortest path $p_{\ell}$ for some $\ell \in \{1, \dotsc, k\}$, the corresponding BOSP instance $\ksppBOSP{p_{\ell}}$ is defined as in \Cref{def:kspp:BOSPinstance} but the source node in $\ksppBOSP{p_{\ell}}$ is not always the actual source node $s$ of our \kspp{k} instance. For $\ksppBOSP{p_{\ell}}$ the source node is $\mathtt{source}(p_{\ell})$. Despite being important for the correctness of \Cref{algo:ksppNew}, this makes sense because in \Cref{def:kspp:BOSPinstance} we assume a shortest path to be given. As discussed in the previous section, the $\ell$\ts{th} shortest path $p_{\ell}$ in the original graph $D$ is not a shortest $s$-$t$-path but its suffix $\subpath{p_{\ell}}{\mathtt{source}(p_{\ell})}{t}$ is a shortest path in a modified version of $D$.
\end{remark}

\begin{algorithm}
	\small{
		\SetKwInOut{Input}{Input}
		\SetKwInOut{Output}{Output}
		\Input{\kspp{k} instance \ksppInst{k}}
		\Output{Solution sequence $P = (p_1, \dotsc, p_k)$ of distinct simple $s$-$t$-paths.}
		\BlankLine
		Priority queue of candidate $s$-$t$-paths $C \leftarrow \emptyset$\;
		$p_1 \leftarrow $ shortest $s$-$t$-path in $D$\label{algo:kspp:p1}\;
		Solution sequence $P \leftarrow (p_1)$\label{algo:kspp:p1Storage}\;
		$p_2 \leftarrow$ \bdassp{} solution for $\ksppBOSP{p_1}$\label{algo:kspp:secondShortest}\;
		$(v,w) \leftarrow$ Parent deviation arc of $p_2$\;
		Add $(v,w)$ to $\mathtt{blocked}(p_1)$\label{algo:kspp:blockP2Dev}\;
		Insert $p_2$ to $C$\label{algo:kspp:secondShortestToQueue}\;
		\For{$\ell \in \{2, \dotsc, k\}$}{
			$p_{\ell} \leftarrow$ Extract path from $C$ with min. costs\label{algo:kspp:extract}\;
			Add $p_{\ell}$ to the solution sequence $P$\label{algo:kspp:storeSolution}\;
			\lIf{$\ell == k$}{\textbf{break}}
			\BlankLine
			$\subpath{q}{\mathtt{source}(p_{\ell})}{t} \leftarrow$ \bdassp{} solution for $\ksppBOSP{p_{\ell}} = (\bar{D}, \mathtt{source}(p_{\ell}), t, \gamma_{p_{\ell}})$ with $\bar{D} = D \setminus \subpath{p_{\ell}}{s}{\mathtt{dev}(p_{\ell})}$\label{algo:ksppNew:firstDev}\;
			\If{$\subpath{q}{\mathtt{source}(p_{\ell})}{t} \neq NULL$} {
				New $s$-$t$-path $q \leftarrow \subpath{p_{\ell}}{s}{\mathtt{source}(p_{\ell})} \circ \subpath{q}{\mathtt{source}(p_{\ell})}{t}$\label{algo:ksppNew:firstPathFromDev}\;
				$(v', w') \leftarrow $ Parent deviation arc of $q$\label{algo:kspp:firstParentDevArc}\;
				Add $(v', w')$ to $\mathtt{blocked}(p_{\ell})$\label{algo:kspp:blockFirstDevArc}\;
				Insert $q$ into $C$\label{algo:kspp:pathToQueue1}\;
			}
			\BlankLine
			$p \leftarrow$ Parent path of $p_{\ell}$\;
			$\subpath{q}{\mathtt{source}(p)}{t} \leftarrow$ \bdassp{} solution for $\ksppBOSP{p} = (\bar{D}, \mathtt{source}(p), t, \gamma_p)$ with $\bar{D} = D\backslash (\subpath{p}{s}{\mathtt{dev}(p)} \cup \mathtt{blocked}(p))$\label{algo:ksppNew:secondDev}\;
			\If{$\subpath{q}{\mathtt{source}(p)}{t} \neq NULL$} {
				New $s$-$t$-path $q \leftarrow  \subpath{p}{s}{\mathtt{source}(p)} \circ \subpath{q}{\mathtt{source}(p)}{t}$\label{algo:ksppNew:secondPathFromDev}\;
				$(v', w') \leftarrow $ deviation arc of $q$\label{algo:ksppNew:secondParentDevArc}\;
				Add $(v', w')$ to $\mathtt{blocked}(p)$\;
				Insert $q$ into $C$\;
			}
		}
		\Return{$P$}\;
		\caption{New algorithm for the \kspp{k} problem.}\label{algo:ksppNew}}
\end{algorithm}

The global data structures of the algorithm are the solution sequence $P$ and a priority queue $C$ of $s$-$t$-paths sorted according to the paths' costs. Both structures are initially empty.
In its initialization phase, the algorithm computes a shortest $s$-$t$-path $p_1$ in $D$ w.r.t $c$ and stores it in $P$ as the first solution in the solution sequence (\Cref{algo:kspp:p1} and \Cref{algo:kspp:p1Storage}). Additionally, a second shortest path $p_2$ is computed applying the \bdassp{} to the $\ksppBOSP{p_1}$ instance (\Cref{algo:kspp:secondShortest}). The obtained path is inserted into $C$ (\Cref{algo:kspp:secondShortestToQueue}). By \Cref{lemma:secondShortestIsChildOfShortest} $p_1$ is the parent path of $p_2$. Every path in $P$ has a list of blocked arcs associated with it. For a path $p$, the list $\mathtt{blocked}(p)$ contains the deviation arcs from $p$'s children paths that are already computed. When looking for further deviations from $p$, we delete the arcs in $\mathtt{blocked}(p)$ from the digraph to ensure that the deviations leading to the already computed children paths of $p$ are not computed again. Thus, since $p_2$ is a child path of $p_1$, the $p_2$'s deviation arc is added to $\mathtt{blocked}(p_1)$ (\Cref{algo:kspp:blockP2Dev}).

After the initialization, the main loop of the algorithm with $k-1$ iterations starts. Every iteration $\ell \in [2, k]$ starts with the extraction of a minimal path from $C$ (\Cref{algo:kspp:extract}), which we call $p_{\ell}$. $p_{\ell}$ is immediately added to $P$ after its extraction and it becomes part of the final solution sequence (\Cref{algo:kspp:storeSolution}). 

\paragraph{First \bdassp{} calculation} Let $(v,w) = (\mathtt{dev}(p_{\ell}), \mathtt{source}(p_{\ell}))$ be the deviation arc from $p_{\ell}$ as defined in \Cref{def:kspp:deviations}. Then, we build the instance $\ksppBOSP{p_{\ell}} = (\bar{D}, w, t, \gamma_{p_{\ell}})$ with $\bar{D} = D \setminus \subpath{p_{\ell}}{s}{v}$. Recall that by \Cref{lemma:kssp:secondShortestShortest}, the suffix $\subpath{p_{\ell}}{w}{t}$ is a shortest $w$-$t$-path in $\bar{D}$. Using the \bdassp{}, a second shortest $w$-$t$-path $\subpath{q}{w}{t}$ in $\bar{D}$ w.r.t. $\gamma_{p_{\ell}}$ is searched. The result, if it exists, is a new suffix for the prefix $\subpath{p_{\ell}}{s}{w}$. Together, both subpaths build a new candidate $s$-$t$-path $q := \subpath{p_{\ell}}{s}{w} \circ \subpath{q}{w}{t}$.

\paragraph{Postprocessing} If $q$ is successfully built, $p_{\ell}$ is its parent path (see \Cref{rem:kssp:parentPathsInAlgo}). Moreover, $q$'s deviation arc is added to the list $\mathtt{blocked}(p_{\ell})$. Finally, $q$ is added to $C$.

\paragraph{Second \bdassp{} calculation} The second \bdassp{} query (\Cref{algo:ksppNew:secondDev}) in every iteration looks for the next-cheapest deviation from the parent path $p$ of the extracted path $p_{\ell}$. When building the corresponding \kspp{2} instance $\ksppBOSP{p}$, the deviation arcs from $p$'s children paths must be deleted from the digraph $D$. Otherwise, the solution to $\ksppBOSP{p}$ would be an already computed deviation. Apart from deleting the arcs in $\mathtt{blocked}(p)$ from $D$, we again delete $p$'s prefix $\subpath{p}{s}{\mathtt{dev}(p)}$ from $p$. This ensures that after the \bdassp{} computation, the concatenation of $\subpath{p}{s}{\mathtt{source}(p)}$, where $w$ is the adjacent node to $v$ in $p$, and the result $\subpath{q}{w}{t}$ is a simple path. If $q$ is successfully built, the algorithm repeats the postprocessing of the first \bdassp{} computation. This query search for the cheapest simple path alternative for $\subpath{p}{w}{t}$ without considering the alternatives that have already been computed.

\subsection{Correctness and Complexity}

In this subsection we sketch the correctness proof and the complexity of \Cref{algo:ksppNew}. The correctness of \Cref{algo:ksppNew} using a black box algorithm to solve the arising \kspp{2} instances is discussed in \citet{Roditty12}.


In \Cref{algo:ksppNew} we use the parent-child relationship of paths introduced in \Cref{def:kspp:deviations}. Formally we would need a proof to show that indeed the computed paths and our usage of this notion in the algorithm are in accordance with the original definition. The following remark gives a strong intuition. The proof can then easily be concluded with an induction step.

\begin{remark}
	\label{rem:kssp:parentPathsInAlgo}
	We know from \Cref{lemma:secondShortestIsChildOfShortest} that $p_2$'s parent path is $p_1$. In the first iteration of \Cref{algo:ksppNew}, we thus build a $\mathtt{source}(p_2)$-$t$-path $\subpath{q}{\mathtt{source}(p_2)}{t}$ in \Cref{algo:ksppNew:firstDev}. It is used to build an $s$-$t$-path $q := \subpath{p_2}{s}{\mathtt{source}(p_2)} \circ \subpath{q}{\mathtt{source}(p_2)}{t}$. Since by definition $\mathtt{source}(p_2) \in p_2$  and $\mathtt{source}(p_2) \notin p_1$, $q$ shares a prefix of maximum length with $p_2$. Hence, $p_2$ induced the BOSP instance $\ksppBOSP{p_2}$ that led to $q$'s computation and $p_2$ is $q$'s parent path.
	
	In the second \bdassp{} query in the first iteration, an $s$-$t$-path $q$ is computed in \Cref{algo:ksppNew:secondDev}. $q$ already starts at $s$ because $s = \mathtt{source}(p_1)$ and thus $s$ is the source node in $\ksppBOSP{p_1}$ (cf. \Cref{kspp:rem:SourceNodes}).    
	In the corresponding digraph, the arcs in $\mathtt{blocked}(p_1)$ are deleted. At  this stage, the list only contains $p_2$'s deviation arc $(v,w) = (\mathtt{dev}(p_2), \mathtt{source}(p_2))$ that was added to the list in \Cref{algo:kspp:blockP2Dev}. Hence, if $\subpath{q}{s}{\mathtt{dev}(q)}$ coincides with $p_1$ until a node $\mathtt{dev}(q)$ that comes after $\mathtt{dev}(p_2)$ along $p_1$, $q$ shares a prefix of maximum length with $p_1$. Otherwise, if $\mathtt{dev}(q)$ does not come after $\mathtt{dev}(p_2)$ along $p_1$, $q$ shares a prefix of maximum length with $p_1$ and $p_2$. By definition, the parent path of $q$ is then set to be the first of these two paths in $P$, i.e., $p_1$. 
	
	Recall that a child's deviation node does not come before its parent's deviation node as remarked already in \Cref{sec:optStructure}. Then, we repeat the arguments from the last paragraphs for any path extracted from $C$ in \Cref{algo:kspp:extract} of \Cref{algo:ksppNew} to proof that the notions from \Cref{def:kspp:deviations} are correctly used in \Cref{algo:ksppNew}.
\end{remark}

The \Cref{algo:ksppNew} requires the deletion of prefixes from the digraph to ensure that it can generate distinct paths when concatenating the suffixes build by the \bdassp{} with the corresponding prefix in the parent path (see \Cref{lemma:kspp:distinct}). Moreover, deleting the prefixes in the graphs $\bar{D}$ used by the \bdassp{} ensures that prefix nodes do not appear in the paths obtained from the \bdassp{}.

\begin{lemma}
	Let $p$ be an $s$-$t$-path in the solution sequence $P$ of \Cref{algo:ksppNew} with deviation node $v$ and deviation arc $(v,w)$. Let $\subpath{q}{w}{t}$ be a second shortest path computed in \Cref{algo:ksppNew:firstDev} or in \Cref{algo:ksppNew:secondDev}. The $s$-$t$-path $q = \subpath{p}{s}{w} \circ \subpath{q}{w}{t}$ is simple.
\end{lemma}

\begin{proof}
	For the computation of $\subpath{q}{w}{t}$, we delete the prefix $\subpath{p}{s}{v}$ from $D$ to build $\bar{D}$. Hence, both subpaths are node-disjoint. As discussed already, the non-negativity of every $\gamma$ ensures that the path output by the \bdassp{} is simple. Hence, $q$ does not contain a cycle.
\end{proof}

The deletion of prefixes and \texttt{blocked} arcs in the graphs $\bar{D}$ in \Cref{algo:ksppNew:firstDev} and in \Cref{algo:ksppNew:secondDev} ensures that every $s$-$t$-path found in \Cref{algo:ksppNew:firstPathFromDev} or in \Cref{algo:ksppNew:secondPathFromDev} of \Cref{algo:ksppNew} is built and added to $C$ only once. This is a property of the deviation tree that partitions the set of $s$-$t$-paths in disjoint sets. 

\begin{lemma}[\citet{Roditty12}, Lemma 3.3.]
	\label{lemma:kspp:distinct}
	Every $s$-$t$-path added to $C$ is only added once.
\end{lemma}

The final correctness statement for \Cref{algo:ksppNew} is proven by induction and uses the correctness of the \bdassp{} and the last two lemmas in this section.

\begin{theorem}[\citet{Roditty12}, Lemma 3.4.]
	\Cref{algo:ksppNew} solves the \kspp{k} problem.
\end{theorem}

We end this section analyzing the running time bound and the memory consumption of \Cref{algo:ksppNew}.

\begin{theorem}
	\label{thm:kspp:RunningTime}
	\Cref{algo:ksppNew} solves a \kspp{k} instance \ksppInst{k} in time
	\begin{equation}
		\label{eq:kspp:runningTime}
		\bigoh{kn \big( n \log{n} + m\big)}.
	\end{equation} 
\end{theorem}

\begin{proof}
	The main loop of \Cref{algo:ksppNew} does $k-1$ iterations. Except in the last iteration where it does not compute new paths, it performs two \bdassp{} computations per iteration. Thus, it computes $(2k - 4) \in \bigoh{k}$ new paths using the \bdassp{}. Using the running time bound for the \bdassp{} derived in \Cref{thm:kspp:bdasspRunningTime}, we obtain the running time bound \eqref{eq:kspp:runningTime} for \Cref{algo:ksppNew}. Thereby we can neglect the effort for the concatenation of paths in \Cref{algo:ksppNew:firstPathFromDev} and in \Cref{algo:ksppNew:firstPathFromDev} since they can be done in $\mathcal{O}(n)$ given that simple paths have at most $(n-1)$ arcs. Moreover, the priority queue operations on $C$ can also be neglected since the queue contains $\mathcal{O}(k)$ elements and we can assume input values of $k$ s.t. $\bigoh{k \log{k}} \subset \bigoh{km}$.
\end{proof}

\begin{theorem}
	\Cref{algo:ksppNew} uses
	$
	\bigoh{kn + n^2 + m}
	$
	memory.
\end{theorem}

\begin{proof}
	By \Cref{thm:kspp:bdasspSpace} we know that any \bdassp{} query in \Cref{algo:ksppNew:firstDev} or in \Cref{algo:ksppNew:secondDev} requires $\bigoh{n^2 + m}$ space. \Cref{algo:ksppNew} does not run multiple \bdassp{} queries simultaneously. We store the paths in the solution sequence $P$, using the deviation tree $T_P$ of $P$ (cf \Cref{def:kspp:deviations}) that allows us to use the parent-child relationships between paths and the notion of deviation arcs, deviation nodes, and source nodes. In $T_P$ every node $v \in V$ can appear multiple times, one per path in $P$. Since simple paths have at most $n-1$ arcs, this results in $\bigoh{kn}$ space.
\end{proof}

\subsection{Implementation details}
\label{sec:kspp:implementationDetails}
The performance of \Cref{algo:ksppNew} in practice depends on the number of iterations required by the \bdassp{} queries. Intuitively, we hope that the search deviates from and returns to the path input to the algorithm after only a few iterations. On big graphs, finding $k$ simple paths between the input nodes $s$ and $t$ is most often a \emph{local} search since only a rather small number of nodes needs to be explored. However, a \bdassp{} query that deviates from the input path but does not return to it fast resembles a One-to-All BOSP algorithm. Thus, it performs a rather \emph{global} search that requires a lot of time. 

The behavior defined above happens mainly when the target node $t$ is not reachable from the source node $\mathtt{source}(p)$ of a \bdassp{} query defined based on an $s$-$t$-path $p$. More precisely, there is always a $\mathtt{source}(p)$-$t$-path in the considered digraph, namely the subpath $\subpath{p}{\mathtt{source}(p)}{t}$ but we are interested in a second shortest $\mathtt{source}(p)$-$t$-path. However, if $p$'s suffix is the only path, the \bdassp{} does not terminate until it empties its priority queue at the beginning of an iteration. This behavior motivates the following \emph{pruning technique}.

\paragraph{\bdassp{} Pruning Using the Paths' Queue}
As soon as \Cref{algo:ksppNew} has at least $k$ $s$-$t$-paths in $P$ and in $C$, i.e., as soon as $|P| + |C| \geq k$, we can possibly end \bdassp{} queries before $t$ is reached or before the \bdassp{} heap becomes empty. In this scenario, set cost $\bar{c} := \max_{p \in C} c(p)$. If the \bdassp{} extracts an $s$-$v$-path $q$ for any $v \in V$ with $c(q) \geq \bar{c}$, the lexicographic ordering of the extracted paths during the \bdassp{} guarantees that no $s$-$t$-path $p$ with costs $c(p) < \bar{c}$ can be build using the suffix computed in that query. Hence, the \bdassp{} query can be aborted. Note that the condition $|P| + |C| \geq k$ is met after the $\frac{k}{2}$\ts{th} iteration at the earliest because in every iteration \Cref{algo:ksppNew} generates at most $2$ new $s$-$t$-paths.

\paragraph{Pruning by Min Paths' Queue Costs}
In graphs with multiple cost equivalent $s$-$t$-paths we may avoid some \bdassp{} queries. Suppose $c^*$ is the minimum cost of paths stored in $C$ at the beginning of an iteration, i.e. $c^* = \min_{p \in C}\{c(p)\}$. We denote the set of paths in $C$ with costs $c^*$ by $C^* \subseteq C$. If at the beginning of an iteration in \Cref{algo:ksppNew} we have $|P| + |C^*| \geq k$, we can terminate the algorithm after extracting the first $k - |P|$ paths from the priority queue and storing them in $P$. The avoided \bdassp{} queries would yield $s$-$t$-paths $p$ with $c(p) \geq c^*$ and thus would not destroy the optimality of the output sequence $P$. 

\section{Experiments}
\label{sec:kspp:experiments}

We now return to 
\Cref{algo:ksppNew} and assess its practical performance by comparing it to the current state of the art: the KM algorithm introduced in \citet{Kurz2016}.

\subsection{Benchmark Setup}
We benchmark \Cref{algo:ksppNew} on $100\times 100$ grid graphs and on road networks from parts of the USA. The choice of an artificially generated set of graphs such as grid graphs and the well known USA road networks from \citep{Dimacs} is common in the \kspp{k} literature.

\paragraph{Grid Graphs} We consider a $100 \times 100$ undirected grid graph and model it as a directed graph $D$ with every edge substituted by two directed arcs as usual. On the digraph $D$ that has $10000$ nodes and $39600$ arcs, we define $10$ different scalar arc cost functions $c$. The arc costs are chosen uniformly and at random between $0$ and $10$. Each of these cost functions, paired with the grid graph, builds a pair $(D,c)$. For each of these pairs, we define $200$ $s$-$t$-pairs, where $s$ and $t$ are chosen uniformly at random from the set of nodes in $D$. Finally, for every tupel $(D, s, t, c)$, we define a \kspp{k} instance \ksppInst{k} using different values for $k$ as shown in \Cref{tab:kspp:gridResults}.

\paragraph{Road Networks} We consider a subset of the USA road networks $D$ included in \citet{Dimacs}. The size of the considered networks as well as their names are in \Cref{tab:kssp:roadInstances}. The cost for an arc in the graph corresponds to the distance between its end nodes. We refer to the resulting arc cost function by $c$. For the $s$-$t$-pairs we draw $200$ $s$-$t$-pairs uniformly and at random from each graph's nodes' set. The final \kspp{k} instances are then defined using different values for $k$ for every tuple $(D,s,t,c)$ as shown in \Cref{tab:kspp:roadResults}.

\begin{table}
	\small
	\centering
	\caption{USA road networks used for experiments. For every graph, we define $200$ $s$-$t$-pairs uniformly and at random. }\label{tab:kssp:roadInstances}
	\begin{tabular}{@{}lSS@{}}
		Road Network & {Nodes}   & {Arcs}  \\
		\midrule
		NY            & 264346  & 733846   \\
		BAY           & 321270  & 800172   \\
		COL           & 43566  & 1057066  \\
		FLA           & 1070376 & 2712798 \\
		LKS           & 2758119 & 6885658 \\
		CTR           & 14081816 & 34292496 \\
		\bottomrule
	\end{tabular}
\end{table}

\paragraph{Benchmark Algorithm}
We compare our implementation of \Cref{algo:ksppNew} that is available in \citep{code} with the implementation of the KM algorithm \citep{Kurz2016} kindly provided to us by the authors. Both algorithms are implemented in C++ and use the same datastructures to store the graph. We explained the choice of the KM algorithm for our benchmarks already in \Cref{sec:kspp:intro}.

\paragraph{Environment}
We used a computer with an Intel(R)~Xeon(R)~Gold~6338 processor and assigned \num{30}\unit{GB} of RAM and \num{2}\unit{h}=\num{7200}\unit{s} for each instance. Both algorithms are compiled using the g++ compiler and the -O3 compiler optimization flag. Our code repository \citep{code} includes the scripts used to run the KM algorithm (even though the code itself needs to be requested from the authors). This is relevant since the implementation includes some optional arguments that highly impact its performance. Our chosen configuration resembles the performance of the best version of the algorithm referenced in \citet{Kurz2016}.
\subsection{Results}

To mitigate the impact of outliers on the reported averages we always report geometric means in this section. In \Cref{tab:kspp:geoMeanDetails} we specify the format of the columns used in the tables in this section. We used the publicly available files \texttt{results/evaluationGrids.ipynb} and \texttt{results/evaluationRoad.ipynb} in \citep{code} to generate the tables and figures. The corresponding \texttt{results} folder also contains the detailed output lines for every solved instance. Note that for every row in \Cref{tab:kspp:gridResults} and in \Cref{tab:kspp:roadResults} the evaluation scripts automatically generate scatter plots like the ones in \Cref{fig:ny-100000} - \Cref{fig:lks-1000000}.

\begin{table}
	\small
	\centering
	\caption{Details on the report of geometric means in our tables in this section.}
	\label{tab:kspp:geoMeanDetails}
	\begin{tabular}{l c l}
		Columns & Unit & Accuracy \\
		\midrule
		Time & \unit{s} & Hundreds \\
		Speedup = $T_{KM}/T_{NA}$ & $\backslash$ & Hundreds \\
		Iterations & amount & Integer \\
		Trees and \bdassp{} & amount & Integer \\
		\bottomrule
	\end{tabular}
\end{table}

Speedups are calculated as the time needed by the KM algorithm divided by the time needed by \Cref{algo:ksppNew}. Thus, speedups greater than $1$ indicate a faster running time for \Cref{algo:ksppNew}. Instances that were not solved by any of the two algorithms are not included in our reports. If an instance was solved by one of the algorithms only, we assume a running time of $T = \num{2}\unit{h}=\num{7200}\unit{s}$ for the other algorithm. 

\subsection{Grid Graphs}
We summarize our results on grid graphs in \Cref{tab:kspp:gridResults}. \Cref{tab:kspp:explanations} explains the column names that are not self-explanatory. For the chosen \kspp{k} instances on grids, we end up considering $2000$ instances for every fixed value of $k$. As shown in \Cref{tab:kspp:gridResults} up to $k = 10^5$, the KM algorithm and \Cref{algo:ksppNew} solved all instances. For $k = 5\times 10^5$ the KM algorithm fails to solve $49$ instances and for $k = 10^6$ it does not solve $445$ instances. 
All instances that are not solved by the KM algorithm are due to the memory limit. 
In contrast, \Cref{algo:ksppNew} manages to solve all instances for every value of $k$.
Regarding the speedup, we observe that \Cref{algo:ksppNew} consistently outperfomrs the KM algorithm. Moreover, the speedup increases as $k$ increases. For $k \geq 5\times 10^4$ the speedup is close to or higher than an order of magnitude.

The reason for both the unsolved instances and the slower running times of the KM algorithm is that the algorithm is forced to compute too many shortest path trees as shown in the column \emph{Trees} in \Cref{tab:kspp:gridResults}. In fact, for $k \geq 10^4$ it approximately needs to compute a tree for every $5$\ts{th} solution path. This is because the considered grid graph is originally an undirected graph. After converting it to a directed graph by adding antiparallel arcs, it contains many cycles. The KM algorithm initially computes a shortest path tree and it can build $\bigoh{m}$ paths from that tree by switching tree arcs and non-tree arcs. This procedure works as long as the switch does not cause the \emph{next} $s$-$t$-path to be non-simple. Given the great amount of antiparallel arcs in the considered grid graph, the KM algorithm cannot build many simple paths from one shortest path tree. 

The good performance of \Cref{algo:ksppNew} on grid graphs is due to the low number of iterations that it requires in every \bdassp{} search. The column \emph{\bdassp{}} in \Cref{tab:kspp:gridResults} reports how many out of at most $2k$ \bdassp{} queries are performed on average. We see that due to the \emph{Pruning by Min Paths' Queue Costs} described in \Cref{sec:kspp:implementationDetails}, \Cref{algo:ksppNew} can skip around $20\%$ of the \bdassp{} queries on average. Whenever the conducted queries find a new path, the average number of iterations in every \bdassp{} query ranges from $20$ to $28$ as shown in the column \emph{Iterations \cmark}. This means that the computed second simple shortest paths in \Cref{algo:ksppNew:firstDev} and in \Cref{algo:ksppNew:secondDev} are found fast. Moreover, in column \emph{\bdassp{} \xmark} we report the number of \bdassp{} queries that do not find a suitable suffix to build a new $s$-$t$-path. There \bdassp{} queries can fail either because $t$ is not reachable or because the \emph{\bdassp{} Pruning using the Paths' Queue} explained in \Cref{sec:kspp:implementationDetails} avoids the computation of the suffix. As reported in the column \emph{Iterations \xmark}, the stopping condition is fulfilled after $16$ to $17$ \bdassp{} iterations hence avoiding the computation of unneeded and large sets of efficient paths. 
\begin{landscape}
	\begin{table}
		\caption{Explanation of columns in \Cref{tab:kspp:gridResults}, \Cref{tab:kspp:roadResults}}
		\label{tab:kspp:explanations}
		\centering
		\begin{tabular}{c l l}
			Algorithm                               & Column Name       & Explanation \\
			\toprule
			\toprule
			KM                                     & Trees             & Shortest path trees computed on average.\\
			\midrule
			\multirow{4}{*}{\Cref{algo:ksppNew}}   & \bdassp{}         & \bdassp{} queries on average. At most $2k - 4$.\\
			& \bdassp{} \xmark  & Average number of \bdassp{} queries that did not reach $t$.\\
			& Iterations \cmark & Avg. iterations in \bdassp{} queries that reached $t$.\\
			& Iterations \xmark & Avg. iterations in \bdassp{} queries that did not reach $t$.\\
		\end{tabular}
	\end{table}
	
	\begin{longtable}{r r r r r r r r r r r}
		\caption{Summarized results obtained from the \kspp{k} instances defined on grid graphs. The column \emph{\bdassp{} \xmark} reports the number of \bdassp{} runs that did not return a second shortest path. If \emph{Iterations \xmark} is a small number, the non-existence of a relevant second shortest path could be proven fast (cf. \Cref{sec:kspp:implementationDetails}).}\label{tab:kspp:gridResults}\\
		
		\multirow{2}{*}{$k$}   &	\multicolumn{3}{c}{KM}		&	\multicolumn{6}{c}{\Cref{algo:ksppNew}} 														& \multirow{2}{*}{SPEEDUP}\\
		\cmidrule(lr){2-4} 					\cmidrule(lr){5-10}
		& Solved  & Trees & Time 	& Solved & \bdassp{} & \bdassp{} \xmark & Iterations \cmark & Iterations \xmark & Time & \\
		\midrule
		1000 &      2000 &        97 &      0.05 &      2000 &      1604 &       245 &                  28 &                  17 &      0.01 &    3.68\\
		5000 &      2000 &       778 &      0.25 &      2000 &      8182 &      1303 &                  25 &                  17 &      0.05 &    4.66\\
		10000 &      2000 &      1847 &      0.53 &      2000 &     16387 &      2656 &                  25 &                  16 &      0.10 &    5.17\\
		50000 &      2000 &     11207 &      4.82 &      2000 &     82481 &     13541 &                  23 &                  16 &      0.49 &    9.76\\
		100000 &      2000 &     24011 &      9.95 &      2000 &    167223 &     28760 &                  22 &                  16 &      1.03 &    9.70\\
		500000 &      1951 &    130124 &     65.29 &      2000 &    836001 &    142677 &                  21 &                  16 &      6.16 &   10.61\\
		1000000 &      1555 &    220368 &    249.81 &      2000 &   1681693 &    291944 &                  20 &                  16 &     12.19 &   20.50\\    
		\bottomrule
	\end{longtable}
\end{landscape}

\subsection{Road Networks}

In \Cref{tab:kspp:roadResults} we summarize the results obtained on the road networks. Again, \Cref{tab:kspp:explanations} explains the column names that are not self-explanatory. For every road network and every value of $k$ \Cref{tab:kspp:roadResults} contains a row showing the average results over the $200$ possibly solved instances in that group.

\begin{figure}[b]
	\minipage{0.49\textwidth}
	\includegraphics[width=\linewidth]{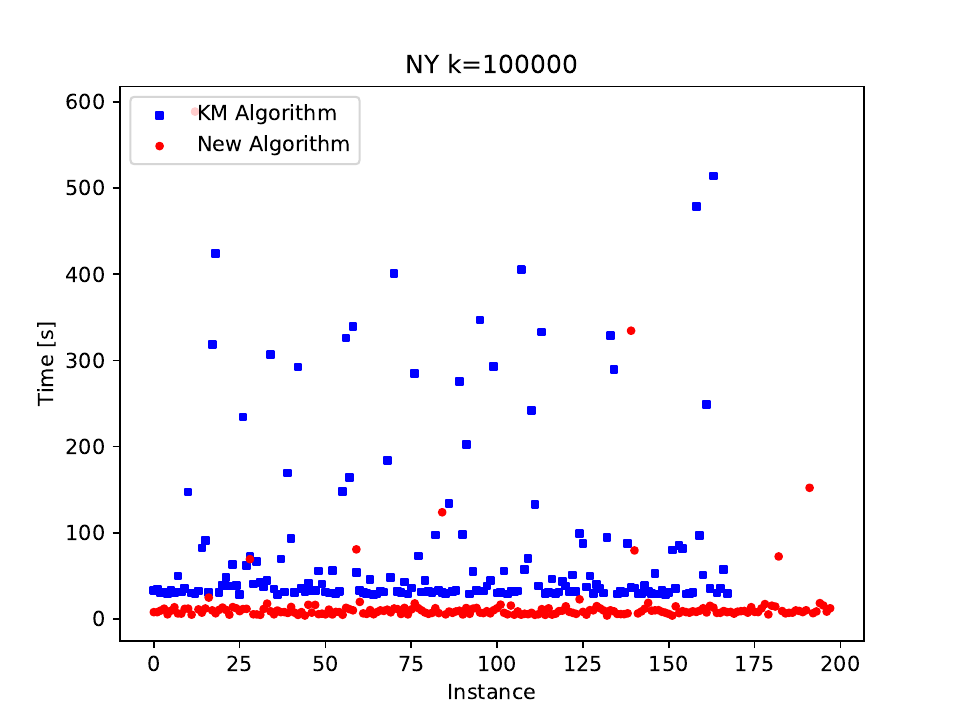}
	\caption{Results obtained from the $200$ instances (if solved) defined on NY networks with $k = 10^5$.}\label{fig:ny-100000}
	\endminipage\hfill
	\minipage{0.49\textwidth}%
	\includegraphics[width=\linewidth]{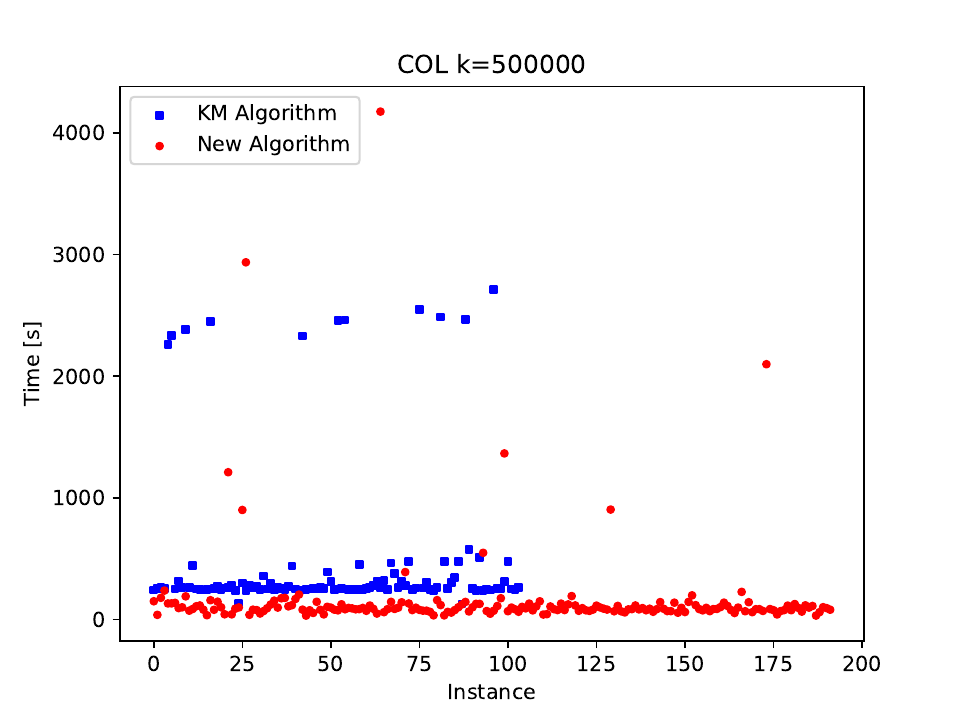}
	\caption{Results obtained from the $200$ instances (if solved) defined on COL networks with $k = 5\times10^5$.}\label{fig:col-500000}
	\endminipage
\end{figure}

\paragraph{Solvability}
The first noticeable difference between both algorithm is that even on the smallest NY network, the KM algorithm fails to solve a considerable amount of instances when $k \geq 5\times 10^4$ (see also \Cref{fig:ny-100000}). Interestingly, also on the much bigger networks, $k = 5\times 10^4$ constitutes a threshold beyond which the KM algorithm struggles to solve multiple instances. \Cref{fig:col-500000} shows an example.
A look at the \emph{KM Time} column unveils that the average running times of the KM algorithm are way below the time limit $T = \num{7200}\unit{s}$. Indeed, the KM algorithm's bottleneck regarding solvability is, as on grid graphs, the memory limit of $\num{30}\unit{GB}$. 
On graphs smaller than LKS, \Cref{algo:ksppNew} manages to solve $\geq \SI{90}{\percent}$ of the instances with $k < 10^6$. The percentage of solved instances with $k \geq 5\times 10^5$ on the LKS and the CTR networks decreases rapidly. 
Again, time is not the problem. Whenever \Cref{algo:ksppNew} fails to solve an instance, its because it hits the memory limit. At this point it is worth noting that instances with $k \geq 10^5$ on road networks are novel in the \kspp{k} literature for algorithms matching the running time bound derived in \Cref{thm:kspp:RunningTime}.

\begin{figure}[b]
	\minipage{0.49\textwidth}
	\includegraphics[width=\linewidth]{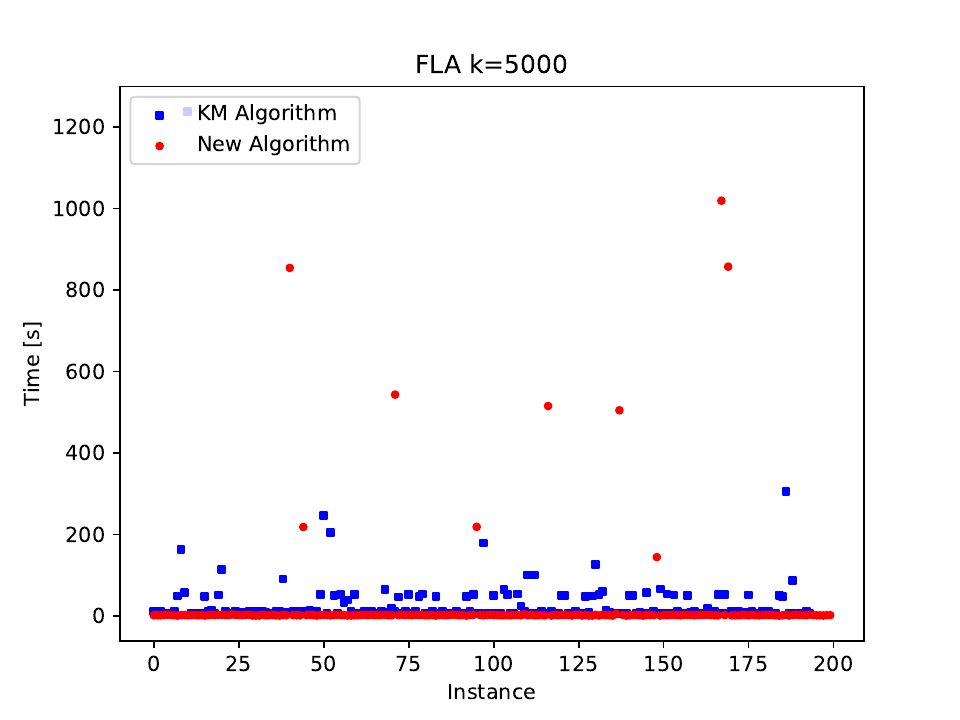}
	\caption{Results obtained from the $200$ instances defined on FLA networks with $k = 5000$.}\label{fig:fla-5000}
	\endminipage\hfill
	\minipage{0.49\textwidth}%
	\includegraphics[width=\linewidth]{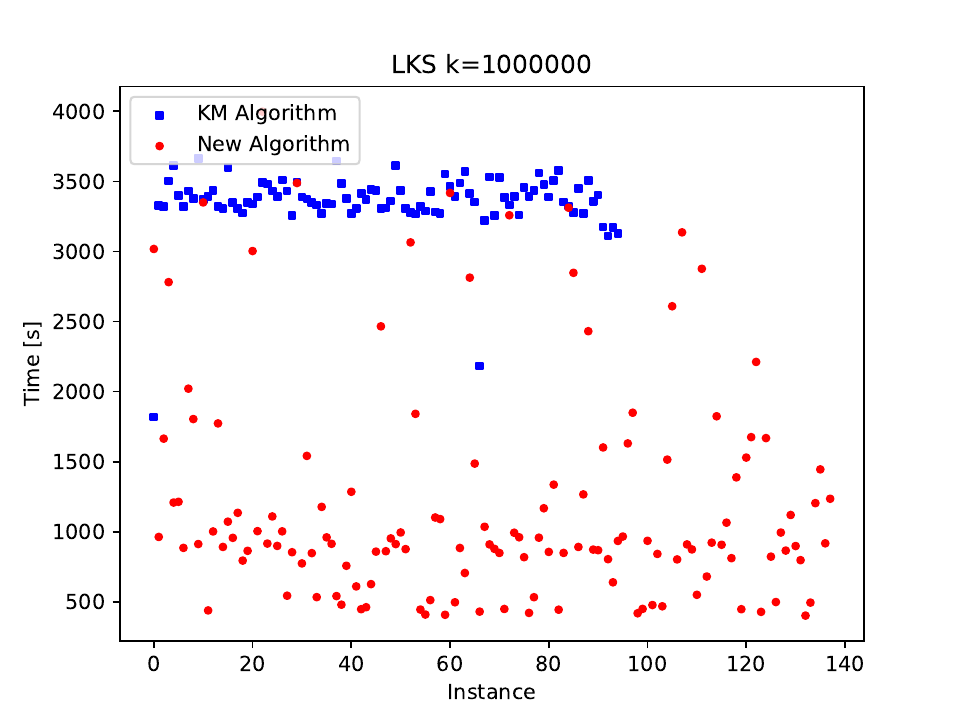}
	\caption{Results obtained from the $200$ instances (if solved) defined on LKS networks with $k = 10^6$.}\label{fig:lks-1000000}
	\endminipage
\end{figure}

\paragraph{Running Times}
For $k < 10$ the KM algorithm is always faster than \Cref{algo:ksppNew}. For every other value of $k$ and for every considered road network, \Cref{algo:ksppNew} is faster on average. We can observe clearly that for every graph, the speedup grows as the value of $k$ grows. The actual values for the speedup favor \Cref{algo:ksppNew} most clearly on the BAY instances. On this graph, speedups of over an order of magnitude on average are reached for $k=5000$ already. The speedup correlates with the number of shortest path computations required by the KM algorithm. The BAY network seems to be particularly hard in this regard. FLA and LKS are interesting networks. On many instances, there are often $k$ $s$-$t$-paths with the cost of a shortest path, regardless of the value of $k$. Additionally, the KM algorithm manages to solve instances on this huge networks computing less than $51$ and less than $10$ shortest path trees in FLA and LKS, respectively. For that reason, the speedup achieved by \Cref{algo:ksppNew} on this graphs is smaller. See \Cref{fig:fla-5000} for a visual example. On such graphs, the main effort by the KM algorithm is checking if the computed paths are simple. Still, for large values of $k$, \Cref{algo:ksppNew} outperforms the KM algorithm on these networks regarding solvability and speed (see \Cref{fig:lks-1000000}).
We can also observe in \Cref{tab:kspp:roadResults} that the pruning techniques discussed in \Cref{sec:kspp:implementationDetails} work well in practice. The column \emph{\bdassp{} \xmark} reports the average number of \bdassp{} queries that do not find a relevant second shortest simple path. These searches, as explained in \Cref{sec:kspp:implementationDetails}, could cause the \bdassp{} queries to compute minimum complete sets of efficient paths for every reachable node. However, using our pruning techniques, we can see in the column \emph{Iterations \xmark} that the average number of iterations on these searches remains low. Often the required iterations on average are even lower than the iterations needed in succesful \bdassp{} queries (see column \emph{Iterations \cmark}).

\section{Conclusion}
We use the black box $k$-Shortest Simple Path (\kspp{k}) algorithm by Roditty and Zwick \citep{Roditty12} to solve the problem. This algorithm solves at most $2k$ instances of the Second Shortest Simple Path (\kspp{2}) problem as a subroutine. In their paper, the authors do not specify how to solve the subroutine efficiently. Since it is a scalar optimization problem, solving it using biobjective path search sounds counter intuitive. However, in this paper we have shown the \kspp{2} can be solved as a biobjective problem using an appropriate biobjective arc cost function. By doing so, we still adhere to the state of the art asymptotic running time bound for the problem. Moreover, we can avoid the nodewise comparison of paths to determine if the computed (sub)paths during our biobjective search are simple; a constant time dominance check suffices. Given a shortest path with $\ell$ nodes, other \kspp{k} algorithms need $\ell$ One-to-One Shortest Path computations to find a second shortest path. Our biobjective approach considers these $\ell$ searches in one biobjective path search and stops as soon as the required second shortest path is found. For these reasons we are able to solve large scale \kspp{k} instances efficiently in practice. Our experiments support this claim.

\begin{landscape}
	\normalfont
	\begin{longtable}{r r r r r r r r r r r}
		\caption{Results obtained from the \kspp{k} instances defined on road networks. If \emph{Iterations \xmark} is a small number, the non-existence of a relevant second shortest path could be proven fast (cf. \Cref{sec:kspp:implementationDetails}).}\label{tab:kspp:roadResults}\\
		\multirow{2}{*}{$k$}		& \multicolumn{3}{c}{KM}		&	\multicolumn{6}{c}{\Cref{algo:ksppNew}} 														& \multirow{2}{*}{SPEEDUP}\\
		\cmidrule(lr){2-4} 					\cmidrule(lr){5-10}
		& Solved  & Trees & Time 	& Solved & \bdassp{} & \bdassp{} \xmark & Iterations \cmark & Iterations \xmark & Time & \\
		\midrule
		
		\multicolumn{11}{c}{NY}\\
		\midrule
		
		10 &       200 &         1 &      0.05 &       200 &        15 &         4 &                 186 &                 130 &      0.10 &    0.52\\
		100 &       200 &         4 &      0.15 &       200 &       188 &        51 &                 140 &                  98 &      0.12 &    1.28\\
		1000 &       200 &        20 &      1.13 &       200 &      1935 &       523 &                 112 &                  88 &      0.23 &    4.98\\
		5000 &       200 &       109 &      4.48 &       200 &      9760 &      2677 &                 100 &                  82 &      0.68 &    6.57\\
		10000 &       197 &       220 &      8.98 &       200 &     19552 &      5363 &                  96 &                  79 &      1.18 &    7.61\\
		50000 &       182 &       987 &     47.34 &       200 &     97824 &     26736 &                  88 &                  76 &      5.20 &    9.11\\
		100000 &       168 &      1643 &    113.67 &       198 &    196277 &     53170 &                  84 &                  69 &      9.73 &   11.68\\
		500000 &       125 &      4405 &    867.08 &       198 &    980103 &    266634 &                  78 &                  66 &     48.72 &   17.80\\
		1000000 &        91 &      4116 &   1975.73 &       196 &   1959033 &    532676 &                  77 &                  61 &     90.82 &   21.75\\
		\midrule
		
		\multicolumn{11}{c}{BAY}\\
		\midrule
		
		10 &       200 &         2 &      0.07 &       200 &        14 &         3 &                 215 &                 168 &      0.10 &    0.71\\
		100 &       200 &         6 &      0.20 &       200 &       189 &        50 &                 182 &                 144 &      0.12 &    1.67\\
		1000 &       200 &        67 &      2.00 &       200 &      1961 &       520 &                 145 &                 134 &      0.30 &    6.70\\
		5000 &       198 &       439 &     11.17 &       200 &      9868 &      2650 &                 128 &                 123 &      0.96 &   11.63\\
		10000 &       190 &       847 &     23.18 &       200 &     19758 &      5303 &                 122 &                 119 &      1.70 &   13.60\\
		50000 &       151 &      3598 &    170.32 &       199 &     98883 &     26579 &                 111 &                 113 &      7.97 &   21.37\\
		100000 &       127 &      5710 &    438.19 &       199 &    197836 &     53240 &                 107 &                 113 &     16.77 &   26.13\\
		500000 &        61 &     12297 &   2528.40 &       190 &    991928 &    263977 &                 104 &                  81 &     77.39 &   32.67\\
		1000000 &        40 &     15572 &   4113.70 &       189 &   1983487 &    526322 &                 101 &                  79 &    138.60 &   29.68\\
		\midrule
		
		\multicolumn{11}{c}{COL}\\
		\midrule
		
		10 &       200 &         1 &      0.09 &       200 &        11 &         3 &                 128 &                 152 &      0.13 &    0.69\\
		100 &       200 &         3 &      0.22 &       200 &       141 &        46 &                 109 &                 133 &      0.15 &    1.41\\
		1000 &       200 &        11 &      1.81 &       200 &      1557 &       420 &                  79 &                 123 &      0.33 &    5.45\\
		5000 &       198 &        38 &      9.61 &       200 &      8300 &      2506 &                  70 &                 120 &      1.10 &    8.76\\
		10000 &       193 &        62 &     16.38 &       200 &     16930 &      4763 &                  70 &                 120 &      2.09 &    7.83\\
		50000 &       156 &        98 &    138.73 &       198 &     86874 &     25251 &                  73 &                 112 &     11.11 &   12.49\\
		100000 &       140 &       117 &    337.81 &       195 &    174875 &     48851 &                  73 &                  97 &     23.33 &   14.48\\
		500000 &       104 &       115 &   1433.21 &       192 &    883570 &    234159 &                  69 &                  92 &    106.66 &   13.44\\
		1000000 &        92 &       124 &   2396.19 &       188 &   1802480 &    459770 &                  72 &                  86 &    202.93 &   11.81\\
		\midrule
		
		\multicolumn{11}{c}{CAL}\\
		\midrule
		
		10 &       200 &         1 &      0.46 &       200 &        11 &         4 &                 258 &                 281 &      0.58 &    0.79\\
		100 &       200 &         2 &      0.83 &       200 &       141 &        45 &                 257 &                 203 &      0.63 &    1.32\\
		1000 &       200 &         5 &      4.20 &       200 &      1680 &       374 &                 223 &                 193 &      1.24 &    3.38\\
		5000 &       197 &        16 &     19.48 &       200 &      9122 &      2374 &                 214 &                 186 &      4.43 &    4.39\\
		10000 &       192 &        25 &     42.71 &       200 &     18463 &      4760 &                 204 &                 178 &      8.56 &    4.99\\
		50000 &       166 &        61 &    317.06 &       197 &     95921 &     24492 &                 185 &                 145 &     47.17 &    6.72\\
		100000 &       152 &        81 &    654.53 &       198 &    192603 &     49266 &                 179 &                 149 &     92.33 &    7.09\\
		500000 &        93 &       208 &   2942.28 &       195 &    964115 &    246741 &                 165 &                 129 &    454.34 &    6.48\\
		1000000 &        82 &       325 &   4287.38 &       178 &   1946894 &    500079 &                 172 &                 124 &    927.41 &    4.62\\
		\midrule
		
		\multicolumn{11}{c}{FLA}\\
		\midrule
		
		10 &       200 &         1 &      0.21 &       200 &        10 &         4 &                 176 &                 162 &      0.32 &    0.66\\
		100 &       200 &         2 &      0.52 &       200 &       132 &        41 &                 147 &                 194 &      0.37 &    1.41\\
		1000 &       200 &         5 &      3.46 &       200 &      1468 &       417 &                 105 &                 198 &      0.80 &    4.34\\
		5000 &       194 &        14 &     17.44 &       200 &      7853 &      2448 &                  78 &                 181 &      2.63 &    6.64\\
		10000 &       188 &        22 &     30.58 &       200 &     15763 &      4858 &                  72 &                 179 &      5.17 &    5.91\\
		50000 &       168 &        51 &    184.12 &       193 &     80003 &     23883 &                  67 &                 113 &     24.91 &    7.39\\
		100000 &       147 &        43 &    505.38 &       192 &    161864 &     47771 &                  72 &                 107 &     42.20 &   11.98\\
		500000 &       114 &        38 &   2190.95 &       187 &    821343 &    229620 &                  85 &                  98 &    220.34 &    9.94\\
		1000000 &        92 &        51 &   3413.61 &       174 &   1699749 &    464313 &                 111 &                  88 &    512.08 &    6.67\\
		\midrule
		\multicolumn{11}{c}{LKS}\\
		\midrule
		
		10 &       200 &         1 &      0.53 &       200 &        10 &         3 &                 172 &                 242 &      0.82 &    0.64\\
		100 &       200 &         1 &      0.99 &       200 &       114 &        45 &                 175 &                 176 &      0.87 &    1.14\\
		1000 &       200 &         2 &      5.25 &       200 &      1242 &       344 &                 139 &                 138 &      1.52 &    3.46\\
		5000 &       198 &         3 &     26.58 &       200 &      6638 &      2102 &                 130 &                 167 &      4.71 &    5.65\\
		10000 &       194 &         4 &     52.56 &       200 &     13596 &      4470 &                 123 &                 176 &      9.25 &    5.68\\
		50000 &       182 &         4 &    317.58 &       199 &     69638 &     23313 &                 102 &                 144 &     49.39 &    6.43\\
		100000 &       172 &         4 &    653.23 &       199 &    141236 &     47241 &                  97 &                 144 &     96.96 &    6.74\\
		500000 &       122 &         5 &   2883.55 &       171 &    735793 &    231196 &                  95 &                 134 &    657.57 &    4.39\\
		1000000 &        95 &         7 &   4692.76 &       138 &   1595516 &    437025 &                  97 &                 131 &   1465.32 &    3.20\\
		\midrule
		\multicolumn{11}{c}{CTR}\\
		\midrule
		10 &       200 &         1 &      4.39 &       200 &         9 &         4 &                 200 &                 285 &      5.42 &    0.81\\
		100 &       200 &         1 &      8.63 &       200 &       105 &        43 &                 194 &                 204 &      5.70 &    1.51\\
		1000 &       200 &         1 &     45.66 &       200 &      1086 &       288 &                 159 &                 153 &      9.42 &    4.85\\
		5000 &       199 &         1 &    213.61 &       199 &      5585 &      1943 &                 142 &                 147 &     32.08 &    6.66\\
		10000 &       198 &         1 &    431.95 &       199 &     11366 &      4135 &                 129 &                 150 &     69.49 &    6.22\\
		50000 &       168 &         2 &   2135.44 &       199 &     59046 &     22443 &                 107 &                 162 &    433.13 &    4.93\\
		100000 &       154 &         2 &   4034.49 &       198 &    119179 &     43441 &                 109 &                 151 &    916.97 &    4.40\\
		\bottomrule
	\end{longtable}
\end{landscape}

\bibliographystyle{unsrtnat}


\end{document}